\documentclass[11pt]{article}
\usepackage{amsmath}
\usepackage{amssymb}
\usepackage{amsthm,amsxtra}
\newlength{\mytopmargin}
\newlength{\myleftmargin}
\newtheorem{lemma}{Lemma}
\newtheorem{prop}[lemma]{Proposition}
\newcommand{\zz}{\mathbb Z}

\begin{document}
\vspace{4cm}
\noindent
{\bf Discrete Painlev\'e equations and random matrix averages}

\vspace{5mm}
\noindent
P.J.~Forrester and N.S.~Witte${}^\dagger$

\noindent
Department of Mathematics and Statistics
${}^\dagger$(and School of Physics),
University of Melbourne, \\
Victoria 3010, Australia ;
email: p.forrester@ms.unimelb.edu.au; n.witte@ms.unimelb.edu.au

\small
\begin{quote}
The $\tau$-function theory of Painlev\'e systems is used to derive
recurrences in the rank $n$ of certain random matrix averages over
$U(n)$. These recurrences involve auxilary quantities which satisfy
discrete Painlev\'e equations. The random matrix averages
include cases which can be interpreted as eigenvalue distributions
at the hard edge and in the bulk of matrix ensembles with unitary
symmetry. The recurrences are illustrated by computing the value of
a sequence of
these distributions as $n$ varies, and demonstrating convergence to
the value of the appropriate limiting distribution.
\end{quote}

\section{Introduction}
\subsection{Motivations and objectives}
In a recent series of papers \cite{FW00}--\cite{FW02c} we have shown how
the Okamoto $\tau$-function theory of Painlev\'e systems can be applied
to rederive known evaluations of certain random matrix averages
in terms of Painlev\'e transcendents. Moreover it was shown how this
theory could similarly be used to evaluate 
random matrix averages not known from previous studies, 
and to also yield recurrences of the
discrete Painlev\'e type for the shift by unity of a parameter or
parameters in the same random matrix averages. Subsequent to our works
\cite{FW00,FW01a} two different major theories --- one on the discrete
Riemann-Hilbert problem due to Borodin \cite{Bo00,Bo02a}, and the
other based on the integrable Toeplitz lattice due to Adler and
van Moerbeke \cite{Av95} --- were applied 
in \cite{Bo02,BB02} and \cite{AvM02} respectively
to also provide
recurrences for random matrix averages with respect to a shift by
unity of a parameter. The averages considered were with respect to the
unitary group $U(n)$,
and the shift performed in the rank $n$ of the matrices. As with
our own work, the average itself is related to an auxilary quantity or
quantities, and it is the latter which satisfy the primary coupled
recurrences.

It is our objective in this work to further develop the Okamoto
$\tau$-function theory as it relates to specifying recurrences
for random matrix averages. Whereas in our earlier works recurrences
were obtained mostly with respect to an otherwise continuous parameter
within the average, in the present work, as with the works by Borodin,
and Adler and van Moerbeke, our attention will be focussed on obtaining
recurrences with respect to the rank of the random matrix and thus
the dimension of the average itself (the averages under consideration
couple only to the eigenvalues of the matrix).

Typical of the results of this paper is the recurrence obtained in our
work \cite{FW00} for the particular PIV $\tau$-function
\begin{eqnarray}\label{1.1}
\tau^{IV}[n](t;\mu) & = & {1 \over C}
\int_{-\infty}^t dx_1 \cdots \int_{-\infty}^t dx_n \,
\prod_{j=1}^n e^{-x_j^2} (t - x_j)^\mu \prod_{1 \le j < k \le n}
(x_k - x_j)^2 \nonumber \\
& =: & \Big \langle \prod_{j=1}^n \chi_{(-\infty,t)}^{(j)} (t-x_j)^\mu
\Big \rangle_{{\rm GUE}_n}.
\end{eqnarray}
Here GUE${}_n$ refers to the probability density function
\begin{equation}\label{1.2}
{1 \over C} \prod_{j=1}^n e^{-x_j^2} \prod_{1 \le j < k \le n}
(x_k - x_j)^2,
\end{equation}
with $C$ denoting the normalization,
realized by the eigenvalues of
Hermitian matrices with certain complex Gaussian entries
(see e.g.~\cite{Fo02}) and $\chi_J^{(j)} =1$ for $x_j \in J$,
$\chi_J^{(j)} =0$ otherwise. From \cite[eqs.~(2.86),(2.4),(2.75)]{FW00} we
have that for an appropriate $C_n$ independent of $t$ (according to (\ref{C})
below $C_n = 2n$),
and with $\tau^{IV}[n] := \tau^{IV}[n](t;\mu)$,
\begin{equation}\label{1.3}
C_n {\tau^{IV}[n+1] \tau^{IV}[n-1] \over (\tau^{IV}[n])^2 } =
2n + (2t - f_0[n] - f_2[n]) f_2[n],
\end{equation}
where $f_0[n], f_2[n]$ satisfy the coupled recurrences
\begin{eqnarray}
f_0[n] + f_0[n-1] & = & 2t - f_2[n] + {2n \over f_2[n]}, \qquad
n=1,2,\dots \label{1.4} \\
f_2[n+1] + f_2[n] & = & 2t - f_0[n] + {2(n + \mu + 1) \over f_0[n]},
\qquad n=0,1,\dots \label{1.5}
\end{eqnarray}
These coupled recurrences were shown to be equivalent to a single second
order difference equation known in the literature as the discrete
Painlev\'e I equation. Specification of $f_0[0], f_2[0], \tau^{IV}[0],
\tau^{IV}[1]$ (see (\ref{2.33a}) below) uniquely determines
$\{ f_0[n] \}_{n=1,2,\dots}, \{ f_2[n] \}_{n=1,2,\dots}$ and
$\{\tau^{IV}[n]\}_{n=2,3,\dots}$. As noted in Section 2 below, the more
general PIV $\tau$-function 
\begin{equation}\label{1.6}
\tau^{IV}[n](t;\mu;\xi)  =  
\Big \langle \prod_{j=1}^n (1 - \xi \chi_{(t,\infty)}^{(j)}) (t-x_j)^\mu
\Big \rangle_{{\rm GUE}_n}
\end{equation}
also satisfies the system (\ref{1.3})--(\ref{1.5}).

Recurrences with respect to the dimension of the random matrix will
also be given for three averages over the unitary group $U(N)$,
known from our earlier work to be $\tau$-functions for certain
Painlev\'e systems. With $z_l := e^{i \theta_l}$ these are
\begin{eqnarray}
\tau^{III'}[N](t;\mu)
& := & \Big \langle \prod_{l=1}^N z_l^\mu
e^{{1 \over 2} \sqrt{t} (z_l + z_l^{-1})} \Big \rangle_{U(N)}
\label{4.1} \\
\tau^{ V}[N](t;\mu,\nu) & := & \Big \langle \prod_{l=1}^N (1 + z_l)^\mu
(1 + 1/z_l)^\nu e^{t z_l} \Big \rangle_{U(N)}
\label{4.2} \\
\tau^{ VI}[N](t;\mu,w_1,w_2;\xi) & := &
\Big \langle \prod_{l=1}^N(1 - \xi \chi_{(\pi - \phi, \pi)}^{(l)})
e^{ w_2 \theta_l} |1 + z_l |^{2 w_1} \Big ( {1 \over t z_l}
\Big )^\mu
( 1 + t z_l)^{2 \mu}
\Big \rangle_{U(N)} \label{4.3},
\end{eqnarray}
where $U(N)$ refers to the probability density function
\begin{equation}
{1 \over (2 \pi )^N N!} \prod_{1 \le j < k \le N} | z_k - z_j |^2,
\qquad (-\pi \le \theta_j \le \pi, \: \: j=1,\dots,N).
\end{equation}
In the case of (\ref{4.1}) we only take the $U(N)$ average as the
definition for $\mu \in \zz$; for general $\mu$ 
the $\tau$-function $\tau^{III'}[N](t;\mu)$ is
to be defined as the Toeplitz determinant given in (\ref{11.1}) below.
Also, as written (\ref{4.2}) is only well defined for $\mu, \nu \in
\zz_{\ge 0}$. However with $z=e^{i \theta}$, use of the identity
\begin{equation}\label{4.3e}
(1 + z)^\mu (1 + 1/z)^\nu = z^{(\mu - \nu)/2} |1 + z|^{\mu+\nu}
\end{equation}
gives
\begin{equation}\label{4.2a}
\tau^V[N](t;\mu,\nu) := \Big \langle \prod_{l=1}^n
 z^{(\mu - \nu)/2}_l |1 + z_l|^{\mu+\nu} e^{t z_l}
\Big \rangle_{U(N)},
\end{equation}
which is well defined for Re$(\mu+\nu) > -1$.

We also indicate how the PV $\tau$-function
\cite{FW01a}
\begin{equation}\label{1.7}
\tilde{\tau}^{V}[n](t;\mu,a;\xi) := 
\Big \langle \prod_{j=1}^n (1 - \xi \chi_{(0,t)}^{(j)}) (x_j - t)^\mu
\Big \rangle_{{\rm LUE}_n}
\end{equation}
and the PVI $\tau$-function \cite{FW02}
\begin{equation}\label{1.8}
\tilde{\tau}^{VI}[n](t;\mu,a,b;\xi) := 
\Big \langle \prod_{j=1}^n (1 - \xi \chi_{(t,1)}^{(j)}) (t-x_j)^\mu
\Big \rangle_{{\rm JUE}_n}
\end{equation}
can be characterized by recurrences.
Here LUE${}_n$ refers to the probability density function 
\begin{equation}\label{1.9}
{1 \over I_N(a)} \prod_{j=1}^n \chi_{(0,\infty)}^{(j)} x_j^a e^{-x_j}
\prod_{1 \le j < k \le n}
(x_k - x_j)^2
\end{equation}
while JUE${}_n$ refers to the probability density function 
\begin{equation}\label{1.10}
{1 \over J_N(a,b)}
\prod_{j=1}^n \chi_{(0,1)}^{(j)} x_j^a (1-x_j)^b 
\prod_{1 \le j < k \le n}
(x_k - x_j)^2.
\end{equation}
The normalizations in (\ref{1.9}) and (\ref{1.10}) are
\begin{equation}\label{I.N}
I_N(a) := \int_0^\infty dx_1 \cdots \int_0^\infty dx_N \,
\prod_{l=1}^N x_l^a e^{-x_l} \prod_{1 \le j < k \le N}
(x_k - x_j)^2 =
\prod_{j=0}^{N-1} \Gamma(2+j) \Gamma(a+1+j)
\end{equation}
and
\begin{eqnarray}\label{J.N}
J_N(a,b) &:= & \int_0^1 dx_1 \, x_1^a(1-x_1)^b \cdots \int_0^1 dx_N \,
x_N^a(1-x_N)^b \prod_{1 \le j < k \le N}(x_k-x_j)^2 \nonumber \\
& = & \prod_{j=0}^{N-1} {\Gamma(a+1+j) \Gamma(b+1+j) \Gamma(2+j) \over
\Gamma(a+b+1+N+j) }.
\end{eqnarray}
We remark that both (\ref{1.9}) and (\ref{1.10}) can be realized as the
eigenvalue probability density function for certain ensembles of random
matrices (see e.g.~\cite{Fo02}). We have not been able to derive
recurrences for (\ref{1.7}) and
(\ref{1.8}) in $n$ only; rather the recurrences to be indicated
also
act on the parameter $a$.

\subsection{Strategy}
The Okamoto theory is based on a Hamiltonian formulation of the
Painlev\'e equations, which in turn can be traced back to Malmquist
\cite{Ma22}. Corresponding to each of the Painlev\'e equations
PII--PVI is a Hamiltonian $H$, which is itself a function of the
conjugate variables $p$ and $q$, the independent variable $t$, and a
number of parameters. The conjugate variables $p$ and $q$ are also 
dependent on the independent variable $t$ and the parameters. By
eliminating $p$ in the Hamilton equations
\begin{equation}\label{5.1}
q' = {\partial H \over \partial p}, \qquad
p' = - {\partial H \over \partial q}
\end{equation}
the Painlev\'e equation in $q$ results, although we have no explicit use
for this defining feature of $H$ below. A particular parameter $n$ is
distinguished and we write $H=H_n$, $p=p_n$, $q=q_n$. Our primary concern
is in so called Schlesinger transformations, which relate the Hamiltonian
system with parameter $n+1$ to the Hamiltonian system with parameter $n$.

One introduces a $\tau$-function $\tau_n$, a function of the
independent variable $t$ and the parameters, by the requirement that
\begin{equation}\label{5.2}
H_n = {d \over dt} \log \tau_n.
\end{equation}
>From the Okamoto theory it is known that
\begin{equation}\label{5.3}
{\tau_{n-1} \tau_{n+1} \over (\tau_n)^2 } =
f(p_n,q_n,t)
\end{equation}
for some explicit polynomial function $f$, typically related to the time
derivative of $H_n$. Furthermore, the Schlesinger transformation theory
gives that $\{p_n,q_n\}$ satisfy coupled first order recurrences
\begin{equation}\label{5.4}
p_{n+1} = g_1(p_n,q_n), \qquad q_{n+1} = g_2(p_n,q_n)
\end{equation}
for some explicit  rational functions $g_1, g_2$. Thus once
$p_0, q_0$ have been specified $\{p_n,q_n\}_{n=1,2,\dots}$ can be
generated from (\ref{5.4}). With this information, and knowledge of
$\tau_0, \tau_1$, (\ref{5.3}) can be iterated to specify
$\{\tau_2,\tau_3,\dots \}$. 

\subsection{Paper outline}
We will devote separate sections to each of the $\tau$-functions
(\ref{1.6})--(\ref{4.3}), with (\ref{1.7}) and (\ref{1.8}) considered
during the discussion of (\ref{4.2}) and (\ref{4.3}) respectively.
In the cases of
(\ref{1.6}),(\ref{4.1}) and (\ref{4.2}) the  Schlesinger transformations
which increment the dimension of respective random matrix averages
are known from our earlier works \cite{FW00,FW01a}. The formulation of the
recurrences is then a straightforward application of the strategy
outlined above. However in the case of 
(\ref{4.3}) there is some complication as one must first change
variables to obtain a random matrix average for which the standard
Schlesinger transformation increments the dimension of the random matrix
average.
In the final section some uses of
our recurrences for the computation of the random matrix averages as they
occur in applied problems will be discussed.

\section{The $\tau$-function sequence $\{\tau^{IV}[n](t;\mu,\xi)\}_{n=0,1,
\dots}$}
\setcounter{equation}{0}
The Hamiltonian for the PIV system is given by \cite{Ok86}
\begin{equation}\label{8.0}
H^{IV} = (2p - q - 2t)pq - 2 \alpha_1 p - \alpha_2 q.
\end{equation}
Let
\begin{equation}\label{8.1}
(\alpha_1,\alpha_2) = (-\mu,-n)
\end{equation}
and write $H^{IV} = H^{IV}_n$ thus distinguishing the parameter
$\alpha_2 = - n$. It was shown in \cite[Prop.~22]{FW00} that corresponding to
the sequence of Hamiltonians $\{H^{IV}_n\}_{n=0,1,\dots}$ is the sequence
of $\tau$-functions $\{\tau^{IV}[n](t;\mu)\}_{n=0,1,
\dots}$ as specified by (\ref{1.1}). Moreover, combining the result of
\cite[Prop.~6]{FW00} with the workings leading to 
\cite[Prop.~7 and Prop,~22]{FW00}
it follows that more generally $\tau^{IV}[n](t;\mu,\xi)$ is a $\tau$-function
for $H_n^{IV}$.

The significance of this latter fact is that with
\begin{equation}\label{8.2}
f_0[n] := 2t + q_n - 2p_n, \qquad f_2[n] := 2p_n
\end{equation}
we know from \cite[eq.~(2.75)]{FW00} that the recurrences 
(\ref{1.3})--(\ref{1.5}) hold, and these recurrences fully determine
$\{\tau^{IV}[n](t;\mu,\xi)\}_{n=2,3,\dots}$ once we specify
$f_0[0], f_1[0]$ in (\ref{1.4}), (\ref{1.5}), and $C_n$,
$\tau^{IV}[0]$, $\tau^{IV}[1]$ in (\ref{1.3}). To determine $C_n$ we require
the fact \cite[eqs.~(2.41), (2.42)]{FW00} that with
\begin{equation}\label{8.3}
C_n = {\gamma_{n+1} \gamma_{n-1} \over \gamma_n^2}, \qquad
\gamma_n e^{t^2 n} \tau^{IV}[n] \mapsto \sigma^{IV}[n], \qquad
\tau^{IV}[0]=\sigma^{IV}[0]=1
\end{equation}
the function $\sigma^{IV}[n]$ has the explicit double Wronskian form
\begin{equation}\label{8.4}
\sigma^{IV}[n] = \det \Big [ {d^{j+k} \over dt^{j+k}} \sigma^{IV}[1]
\Big ]_{j,k=0,\dots,n-1}.
\end{equation}

Now it follows from \cite[eq.~(2.41), Prop.~6]{FW00} that up to a
proportionality constant, which we are free to choose to be unity,
\begin{equation}\label{9.1}
\sigma^{IV}[1] = e^{t^2} \Big ( \int_{-\infty}^\infty - \xi
\int_t^\infty \Big ) (t - x)^\mu e^{-x^2} \, dx.
\end{equation}
Noting that (\ref{9.1}) can be written
$$
\sigma^{IV}[1] = \Big ( \int_{-\infty}^\infty - \xi
\int_0^\infty \Big ) (-x)^\mu e^{-x^2 - 2tx} \, dx
$$
the differentiation required by (\ref{8.4}) becomes simple to perform
and we obtain
\begin{equation}\label{9.1a}
{d^{i+j} \over dt^{i+j}} \sigma^{IV}[1] =
2^{i+j} e^{t^2} \Big ( \int_{-\infty}^\infty - \xi
\int_t^\infty \Big ) (t-x)^{\mu+i+j} e^{-x^2} \, dx.
\end{equation}
Substituting this in (\ref{8.4}) and recalling the workings of 
\cite[proof of Prop.~21]{FW00} we see that
\begin{eqnarray}
\sigma^{IV}[n] & = & {2^{n(n-1)} \over n!} e^{t^2 n}
\Big ( \int_{-\infty}^\infty - \xi
\int_t^\infty \Big ) dx_1 \cdots \Big ( \int_{-\infty}^\infty - \xi
\int_t^\infty \Big ) dx_n \, \prod_{j=1}^n e^{-x_j^2}(t-x_j)^\mu
\nonumber \\
&& \times \prod_{1 \le j < k \le n} (x_k - x_j)^2. \label{9.2}
\end{eqnarray}
It is well known (see e.g.~\cite{Fo02}) that the normalization $C$ in the
definition (\ref{1.2}) of the GUE${}_n$ probability density function has
the explicit form
$$
C = n! 2^{-(n-1)n/2} \pi^{n/2} \prod_{l=0}^{n-1} l!
$$
so (\ref{9.2}) can be written
\begin{eqnarray}
\sigma^{IV}[n] & = & 2^{n(n-1)/2} \pi^{n/2} \prod_{l=0}^{n-1} l!
e^{t^2 n} \Big \langle \prod_{l=1}^n(1 - \xi \chi_{(t,\infty)}^{(l)})
(t - x)^\mu \Big \rangle_{{\rm GUE}_n} \nonumber \\
& = & 2^{n(n-1)/2} \pi^{n/2} \prod_{l=0}^{n-1} l! e^{t^2 n}
\tau^{IV}[n](t;\mu;\xi).
\end{eqnarray}
Recalling (\ref{8.3}) we thus have
\begin{equation}
\gamma_n = 2^{n(n-1)/2} \pi^{n/2} \prod_{l=0}^{n-1} l!
\end{equation}
and this in turn implies
\begin{equation}\label{C}
C_n = 2n.
\end{equation}

Regarding the initial conditions for (\ref{1.4}) and (\ref{1.5}), we require
the facts \cite[proof of Prop.~6]{FW00} that
$$
p_0=0, \qquad q_0 = {d \over dt} \log \tau^{IV}[1].
$$
Thus recalling (\ref{8.2}) and (\ref{1.6}) we have
\begin{equation}\label{10.1}
f_0[0] = 2t + {d \over dt} \log \Big ( \Big ( \int_{-\infty}^\infty - \xi
 \int_t^\infty \Big ) (t - x)^\mu e^{-x^2} \, dx \Big ), \qquad
f_2[0] = 0.
\end{equation}
The initial conditions for (\ref{1.3}) are by definition
\begin{equation}\label{10.2}
\tau^{IV}[0]=1, \qquad \tau^{IV}[1] = {1 \over \sqrt{\pi}}
\Big ( \int_{-\infty}^\infty - \xi
 \int_t^\infty \Big ) (t - x)^\mu e^{-x^2} \, dx.
\end{equation}

In summary, we thus have that the following result holds.

\begin{prop}
Let $\tau^{IV}[n] = \tau^{IV}[n](t;\mu;\xi)$ as specified by (\ref{1.6}). Let
$p_n$, $q_n$ denote the conjugate variables in the Hamiltonian (\ref{8.0})
with parameters given by (\ref{8.1}), and define $f_0[n]$ and $f_2[n]$
in terms of these variables by (\ref{8.2}). We have that
$\{f_0[n]\}_{n=1,2,\dots}$, $\{f_2[n]\}_{n=1,2,\dots}$ and
$\{\tau^{IV}[n]\}_{n=2,3,\dots}$ are determined by the recurrences
(\ref{1.3})--(\ref{1.5}) subject to the initial conditions (\ref{10.1}),
(\ref{10.2}).
\end{prop}

We remark that in the special case $\xi=0$, $\mu \in \zz_{\ge 0}$,
(\ref{1.6}) is a polynomial in $t$, which in view of (\ref{8.3}),
(\ref{8.4}), (\ref{9.1a}) and the integral representation
$$
\int_{-\infty}^\infty (t - ix)^p e^{-x^2} \, dx = \sqrt{\pi} 2^{-p}
H_p(t)
$$
has the explicit form
\begin{equation}\label{10.3}
\tau^{IV}[n] = {1 \over \gamma_n}
\det \Big [ (2i)^{-\mu} i^{-(j+k)}  H_{\mu+j+k}(it)
\Big ]_{j,k=0,\dots,n-1}.
\end{equation}
In this case (\ref{10.1}) and (\ref{10.2}) can be written
\begin{eqnarray}\label{2.33a}
&&f_0[0] = 2t + {2 \mu i H_{\mu - 1}(it) \over H_\mu(it)}, \qquad 
f_2[0] = 0 \nonumber\\
&&\tau^{IV}[0]=1, \qquad \qquad
 \tau^{IV}[1] = { \displaystyle (2i)^{-\mu} \over \gamma_1} H_\mu(it).
\end{eqnarray}
It is of interest to recall the duality formula \cite[eq.~(4.37)]{FW00}
\begin{equation}\label{2.15a}
\tau^{IV}[n](t;\mu,0) = i^{-n\mu} \tau^{IV}[\mu](it;n,0) =
{1 \over \gamma_\mu} \det [ 2^{-n} i^{-(j+k)} H_{n+j+k}(t) ]_{j,k=0,\dots,
\mu-1}
\end{equation}
thus giving (\ref{10.3}) for $n=0,1,\dots$ as a sequence of
$\mu \times \mu$ determinants. 

Another point of interest is that with the initial conditions
(\ref{10.1}) a closed form solution of the coupled recurrences
(\ref{1.4}) and (\ref{1.5}) can be given. Thus it follows from
\cite[eq.~(4.8)]{NY99} that
\begin{equation} 
f_0[n] = 2 {\tau^{IV}[n](t;\mu,\xi) \tau^{IV}[n+1](t;\mu+1,\xi)
\over
\tau^{IV}[n](t;\mu+1,\xi) \tau^{IV}[n+1](t;\mu,\xi)}, \qquad
f_2[n] = n {\tau^{IV}[n+1](t;\mu,\xi) \tau^{IV}[n-1](t;\mu+1,\xi)
\over
\tau^{IV}[n](t;\mu,\xi) \tau^{IV}[n](t;\mu+1,\xi)}
\end{equation}
(the proportionality constants cannot be read off from \cite{NY99};
these are determined by considering the $t \to \infty$ behaviour of
(\ref{1.3})--(\ref{1.5})). 

\section{The $\tau$-function sequence $\{\tau^{III'}[N](t;\mu)\}_{N=0,1,
\dots}$}
\setcounter{equation}{0}
Although (\ref{4.1}) is well defined for all complex $\mu$, only for
$\mu \in \zz$ will we take the $U(N)$ average as the definition. For
general $\mu$ we will make use of a Toeplitz determinant form, obtained
by applying the well known identity
\begin{equation}\label{11.0}
\Big \langle \prod_{l=1}^N w(z_l) \Big \rangle_{U(N)} =
\det \Big [ {1 \over 2 \pi} \int_{-\pi}^\pi w(z) z^{j-k} \, d \theta
\Big ]_{j,k=1,\dots,N}.
\end{equation}
This gives
\begin{eqnarray}\label{11.1}
\Big \langle \prod_{l=1}^N z_l^\mu
e^{{1 \over 2} \sqrt{t} (z_l + z_l^{-1})} \Big \rangle_{U(N)}
& = & \det \Big [ {1 \over 2 \pi} \int_{-\pi}^\pi
e^{i(\mu+j-k)\theta} e^{\sqrt{t} \cos \theta} \, d \theta
\Big ]_{j,k=1,\dots,N} \nonumber \\
& = & \det [I_{\mu + j - k}(\sqrt{t}) ]_{j,k=1,\dots,N}.
\end{eqnarray}
The second equality of (\ref{11.1}) follows from an integral formula
for $I_\nu(z)$, valid for $\nu \in \zz$. We take this latter determinant
as the meaning of $\tau^{III'}[N]$ for general $\mu$.

Now the Hamiltonian for the PIII${}'$ system is given by \cite{Ok87a}
\begin{equation}\label{11.2}
t H^{III'} = q^2 p^2 - (q^2 + v_1 q - t)p + {1 \over 2} (v_1 + v_2) q.
\end{equation}
We showed in \cite[Section 4.3]{FW01a} that with
\begin{equation}\label{11.2a}
(v_1,v_2) = (\mu+N,-\mu+N)
\end{equation}
the quantity
\begin{equation}\label{11.3}
t^{-N\mu/2} \det [I_{\mu + j - k}(\sqrt{t}) ]_{j,k=1,\dots,N}
\Big |_{t \mapsto 4t}
\end{equation}
is a $\tau$-function for the corresponding sequence of Hamiltonians
$\{t H_N^{III'} \}_{N=0,1,\dots}$ (it is still valid to call (\ref{4.1})
a $\tau$-function for a PIII${}'$ system as the extra factor
$t^{-N\mu/2}$ is equivalent to the addition of a constant to the
Hamiltonian (\ref{11.2}), which of course does not alter the Hamilton
equations). From the working in \cite{FW02} summarizing the Okamoto
theory of PIII${}'$, we can deduce the following recurrences for (\ref{4.1}).

\begin{prop}\label{p12}
Let $\tau^{III'}[N] = \tau^{III'}[N](t;\mu)$ as given by (\ref{4.1}), and let
$p_N$, $q_N$ denote the conjugate variables in the Hamiltonian (\ref{11.2})
with parameters (\ref{11.2a}). The sequences $\{\tau^{III'}[N]
\}_{N=0,1,\dots}$, $\{p_N\}_{N=0,1,\dots}$, $\{q_N\}_{N=0,1,\dots}$
satisfy the coupled recurrences
\begin{eqnarray}
&& {\tau^{III'}[N+1] \tau^{III'}[N-1] \over (\tau^{III'}[N])^2 }
\Big |_{t \mapsto 4t} = p_N \qquad (N=1,2,\dots) \label{12.1} \\
&& p_{N+1} = {q_{N}^2 \over t} (p_N - 1) - {\mu q_N \over t} +1
\qquad (N=0,1,\dots) \label{12.2} \\
&& q_{N+1} = - {t \over q_N} + {(1+N) t \over q_N(q_N(p_N-1) - \mu) + t}
\qquad (N=0,1,\dots) \label{12.3}
\end{eqnarray}
subject to the initial conditions
\begin{eqnarray}
&& p_0=0, \qquad q_0 = t {d \over dt} \log t^{-\mu/2} I_\mu(2 \sqrt{t})
\label{12.4}
\\ &&
\tau^{III'}[0]=1, \qquad \tau^{III'}[1] = I_\mu(2\sqrt{t}).
\label{12.5}
\end{eqnarray}
\end{prop}

\noindent
Proof. \quad
The working of \cite[proof of Prop.~4.2]{FW01a}, which in turn is based on
\cite{K99}, together with \cite[eqs.~(4.13), (4.20),(4.21)]{FW01a} tell us that
$$
{\tau^{III'}[N+1] \tau^{III'}[N-1] \over (\tau^{III'}[N])^2 }
\Big |_{t \mapsto 4t} = {\partial \over \partial t} (t H_N^{III'}),
$$
so (\ref{12.1}) now follows from (\ref{11.2}). Furthermore, it was shown in
\cite[eqs.~(4.40),(4.41)]{FW01a} that for the Hamiltonian (\ref{11.2})
with parameters $(v_1,v_2) = (v_1^{(0)}+n,v_2^{(0)}+n)$,
\begin{eqnarray}
p_{n+1} & = & {q_n^2 \over t} (p_n - 1) - {q_n \over 2t} (
v_1^{(0)} - v_2^{(0)}) + 1 \label{3.42a} \\
q_{n+1} & = & - {t \over q_n} + {{1 \over 2} (v_1^{(0)} + v_2^{(0)} +2
+2n) t \over q_n(q_n(p_n-1) - {1 \over 2} (v_1^{(0)} - v_2^{(0)})) + t}.
\label{3.42b}
\end{eqnarray}
Setting $v_1^{(0)} = - v_2^{(0)} = \mu$ gives (\ref{12.2}) and (\ref{12.3}).
The initial conditions (\ref{12.4}) follow from 
\cite[proof of Prop.~4.3]{FW01a}, while the initial conditions (\ref{12.5})
are immediate from (\ref{11.1}). \hfill $\square$

\smallskip
It is known (see \cite[Prop.~4.6]{FW01a}) 
that the sequence $\{q_N\}_{N=0,1,\dots}$ satisfies
the particular (alternate) discrete Painlev\'e II equation
\begin{equation}\label{2.23}
{1+N \over q_N q_{N+1} + t} + {N \over q_{N-1} q_N + t} =
{1 \over q_N} - {q_N \over t} + {N-\mu \over t}, \qquad
N=0,1,\dots
\end{equation}
In the special case $\mu=0$ the sequence $\{p_n\}_{n=0,1,\dots}$ itself is
also determined by a particular discrete Painlev\'e II equation. To see
this, note that (\ref{12.2}) with $\mu=0$ gives
\begin{equation}\label{2.24}
q_N^2 = t {1 - p_{N+1} \over 1 - p_N}.
\end{equation}
Setting
\begin{equation}\label{2.25}
q_N = \sqrt{t} {P_{N+1} \over P_N}, \qquad P_0=1
\end{equation}
we see that (\ref{2.24}) has the unique solution
\begin{equation}\label{2.26}
p_N = 1 - P_N^2.
\end{equation}
Making use of this in Proposition \ref{p12} we obtain the following 
recurrence scheme for $\{\tau^{III'}[N](t,0)\}$, first derived by
Borodin \cite{Bo02} from a discrete Riemann-Hilbert approach, and
subsequently obtained by Adler and van Moerbeke \cite{AvM02} from
their theory of the Toeplitz lattice, and by Baik \cite{Ba02} who
used a Riemann-Hilbert approach distinct from that of Borodin.

\begin{prop}\label{p.3}
We have
$$
1 - P_n^2 =
{\tau^{III'}[N+1] \tau^{III'}[N-1] \over (\tau^{III'}[N])^2 }
\Big |_{t \mapsto 4t \atop \mu=0}
$$
where $\{P_n\}_{n=1,2,\dots}$ satisfies the particular discrete Painlev\'e
II equation
$$
P_{n+1} + P_{n-1} = {n P_n \over \sqrt{t} (1-P_n^2 ) }, \qquad
n=1,2,\dots
$$
subject to the initial conditions
$$
P_0 = 1, \qquad P_1 = {I_1(2\sqrt{t}) \over I_0(2\sqrt{t})}.
$$
\end{prop}

Let us show how $q_N$, like $p_N$, can be written in terms of the
$\tau$-functions. Put 
$$
(t^{-N\mu/2} \tau^{III'}[N](t;\mu) )|_{t \mapsto 4t}
=: \tau_N^\mu
$$ 
and denote the corresponding Hamitonian by $H_N^\mu$.
Denote by $T_1$ ($T_2$) the Schlesinger operators with the action on the
parameters $(v_1, v_2) \mapsto (v_1+1, v_2+1)$ ($(v_1, v_2) \mapsto 
(v_1+1, v_2-1)$). It is known \cite{Ok87a} that
\begin{equation}\label{3.42c}
T_1 t H_N^\mu = t H_{N+1}^\mu = t H_N^\mu + q_N(1 - p_N), \qquad
T_2 t H_N^\mu = t H_N^{\mu+1} = tH_N^\mu - q_N p_N
\end{equation}
where $q_N := q_N^\mu$, $p_N := p_N^\mu$. Using (\ref{3.42c}) together with
(\ref{3.42a}), (\ref{3.42b}) the sought formula can be deduced.

\begin{prop}\label{pr.4}
We have
\begin{equation}\label{3.42d}
q_N = (-1)^N
\sqrt{t} {\tau^{III'}[N](4t,\mu) \tau^{III'}[N+1](4t,\mu+1)  \over
\tau^{III'}[N+1](4t,\mu) \tau^{III'}[N](4t,\mu+1) }.
\end{equation}
\end{prop}

\noindent
Proof. \quad We have
$$
t {d \over dt} \log \Big ( {\tau^\mu_{n+1} \tau^{\mu+1}_{n} \over
\tau^{\mu+1}_{n+1} \tau^\mu_{n} } \Big ) = -t (H_{n+1}^{\mu+1} - 
H_{n+1}^\mu) + t(H_n^{\mu+1} - H_n^\mu) = q_{N+1} p_{N+1} - q_N p_N.
$$
According to (\ref{3.42a}), (\ref{3.42b})
$$
q_{N+1} p_{N+1} = - q_N(p_N-1) + v_1 + 1 - {t \over q_N}.
$$
Thus
\begin{eqnarray*}
t {d \over dt} \log \Big ( {\tau^\mu_{n+1} \tau^{\mu+1}_{n} \over
\tau^{\mu+1}_{n+1} \tau^\mu_{n} } \Big ) & = &
- {1 \over q_N} \Big ( 2 q_N^2 p_N - q_N^2 - (v_1 + 1) q_N + t \Big ) \\
& = & - {1 \over q_N} \Big ( {\partial t H^{III'} \over \partial p_N}
- q_N \Big ) \: = \: - t {d \over dt} \log (q_N/t)
\end{eqnarray*}
where to obtain the final equality, use has been made of the first of the
Hamilton equations. This implies (\ref{3.42d}) up to a proportionality
constant. To determine  the proportionality, $c_N$ say, we use the
asymptotic formula \cite[proof of Cor.~4.5]{FW01a}
$$
\det [ I_{j-k+\mu}(\sqrt{t})]_{j,k=0,\dots,n-1}
\mathop{\sim}\limits_{t \to \infty} e^{n \sqrt{t} -(n^2/4) \log t
+O(1)} 
$$
which in light of (\ref{11.1}) and (\ref{12.1}) implies 
\begin{equation}\label{3.18a}
p_N \mathop{\sim}\limits_{t \to \infty} (4t)^{-1/2} \qquad
(N \ne 0)
\end{equation}
while (\ref{3.42d}) (with the proportionality still unknown) implies
\begin{equation}\label{3.18b}
q_N  \mathop{\sim}\limits_{t \to \infty} c_N \sqrt{t}.
\end{equation}
Substituting in (\ref{12.3}) and taking into consideration (\ref{12.4})
implies $c_N = (-1)^N$.
 \hfill $\square$

\section{The $\tau$-function sequence $\{\tau^V[N](t;\mu,\nu)\}$}
\setcounter{equation}{0}
The definition (\ref{4.2a}) of $\tau^V[N](t;\mu,\nu)$ is well defined
for Re$(\mu+\nu) > -1$. This domain can be extended by using (\ref{11.0})
to rewrite (\ref{4.2a}) as a Toeplitz determinant and evaluating the
integral,
\begin{eqnarray}\label{T5.1}
\tau^V[N](t;\mu,\nu) & = & \det \Big [ {1 \over 2 \pi}
\int_{-\pi}^\pi z^{j-k+(\mu - \nu)/2} |1 + z|^{\mu+\nu} e^{t z} \, d \theta
\Big  ]_{j,k=1,\dots,N} \nonumber \\
& = & \det \Big [ {\Gamma(\mu+\nu+1) \over \Gamma(\mu+j-k+1)
\Gamma(\nu-j+k+1)} \, {}_1 F_1 (-\nu+j-k, \mu+1+j-k; - t )
\Big ]_{j,k=1,\dots,N}. \nonumber \\
\end{eqnarray}
Here the integral evaluation, which is well defined for general complex
$\mu$, $\nu$, follows by expanding the exponential in the first Toeplitz
determinant and evaluating the resulting integrals using the formula
\begin{equation}\label{fbo}
{1 \over 2 \pi} \int_{-\pi}^\pi z^{(a-b)/2} |1 + z|^{a+b} \,
d \theta = {\Gamma(a+b+1) \over \Gamma(a+1) \Gamma(b+1)}.
\end{equation}

In \cite[display eq.~below proof of Prop.~3.6]{FW01a},
$\{\tau^V[N](t;\mu,\nu)\}_{N=0,1,\dots}$ has been identified as the
$\tau$-function sequence corresponding to a particular Schlesinger
operator for the PV system. However some technical details of the
derivation given there leads to complication for the present
purposes, which in fact can be avoided by revising some of the workings
in \cite{FW01a}. Let us then undertake such a program.

We will construct a $\tau$-function sequence relating to the Hamiltonian
\cite{Wa98,K99}
\begin{equation}\label{s1.1}
t H^{V^*} = q(q-1)p(p+t) - (v_2 - v_1 + v_3 - v_4) qp + (v_2 - v_1)p
+ (v_1 - v_3)tq
\end{equation}
for which eliminating $p$ in the Hamilton equations shows $1+ 1/(q-1)$
satisfies the PV equation. Since eliminating $p$ does not give the
Painlev\'e equation in $q$ itself, we refer to this as the PV${}^*$ system.
(In \cite{FW01a} we made use of the mapping between the PV and 
 PV${}^*$ systems \cite{Wa98}, which is in fact unnecessary and is
what leads to complications for the present purposes.) Our interest is
in the particular Schlesinger transformation with action on the parameters
\begin{equation}\label{s1.2}
T_0^{-1}(v_1,v_2,v_3,v_4) = (v_1 - {1 \over 4}, v_2  - {1 \over 4},
v_3 - {1 \over 4}, v_4 + {3 \over 4}).
\end{equation}
We know from \cite[eq.~(2.16)]{FW01a} that
\begin{equation}\label{s1.3}
T_0^{-1} H^{V^*} = H^{V^*} 
\Big |_{\mathbf{v} \mapsto T_0^{-1} \mathbf{v}}.
\end{equation}
This motivates introducing the sequence of Hamiltonians
$$
H^{V^*}_n := H^{V^*}_0 \Big |_{\mathbf{v} \mapsto
(v_1^{(0)} - n/4, v_2^{(0)} - n/4, v_3^{(0)} - n/4,
v_4^{(0)} + 3n/4) }
$$
and the corresponding sequence of $\tau$-functions $\tau_n^{V^*}$
specified so that
\begin{equation}\label{s1.4}
H^{V^*}_n = {d \over dt} \log \tau_n^{V^*}.
\end{equation}

Following \cite{Ok87}, the seed initializing the sequence of $\tau$-functions
is a classical solution to the PV${}^*$ system.

\begin{prop}
Let $v_3^{(0)} - v_4^{(0)} = 0$. Then the PV${}^*$ system admits the solution
\begin{equation}\label{s1.5}
q_0 = 1, \quad tH^{V^*}_0 = (v_1^{(0)} - v_3^{(0)}) t, \quad
\tau_0 = e^{(v_1^{(0)} - v_3^{(0)})t}, \quad
p_0 = t {d \over dt} \log \tau_1^{V^*} + (v_3^{(0)} - v_1^{(0)}) t
\end{equation}
where $ e^{-(v_1^{(0)} - v_3^{(0)})t}\tau_1^{V^*}$ 
satisfies the confluent hypergeometric differential
equation
\begin{equation}\label{s2.0}
t y'' + (v_1^{(0)} - v_2^{(0)} +1 + t) y' + (v_1^{(0)} - 
 v_3^{(0)}) y = 0.
\end{equation}
\end{prop}

\noindent
Proof. \quad Direct substitution of $q_0=1$, $v_3^{(0)} - v_4^{(0)} = 0$
into (\ref{s1.1}) gives the stated value of $t H^{V^*}$. The
final equation in (\ref{s1.5}) follows from (\ref{s1.3}), (\ref{s1.1})
and (\ref{s1.4}) which together give
$$
T_0^{-1} t H^{V^*}_0 := t H^{V^*}_1 = t {d \over dt} \log \tau_1^{V^*}
= t H^{V^*}_0 + p_0 = (v_1^{(0)} - v_3^{(0)})t + p_0. 
$$
Now that the final equation in (\ref{s1.5}) is established, (\ref{s2.0})
can be derived from the second of the Hamilton equations (\ref{5.1})
\begin{equation}\label{s2.1}
t p_0' = - {\partial t H^{V^*}_0 \over \partial q }
\Big |_{q_0 = 1 \atop v_3^{(0)} - v_4^{(0)} = 0} = - \Big (
p_0(p_0+t) - (v_2^{(0)} - v_1^{(0)}) p_0 +
(v_1^{(0)} - v_3^{(0)}) t \Big ),
\end{equation}
by substituting the former equation for $p_0$ throughout.
\hfill $\square$

According to \cite[proof of Prop.~2.2]{FW01a}, with
\begin{equation}\label{33.0}
\bar{\tau}_n := t^{n^2/2} e^{(v_4^{(0)} - v_1^{(0)} + n) t}
\tau_n^{V^*}
\end{equation}
and $\tau_0^{V^*}$ as in (\ref{s1.5}) so that $\bar{\tau}_0=1$ (recall that
in (\ref{s1.5}) we require $v_3^{(0)} - v_4^{(0)} = 0$), the sequence
$\{ \bar{\tau}_n \}_{n=2,3,\dots}$ is specified by the determinant formula
\begin{equation}\label{33.1}
\bar{\tau}_n = \det [ \delta^{j+k} \bar{\tau}_1 ]_{j,k=0,\dots,n-1}.
\end{equation}
For an appropriate choice of the solution of (\ref{s2.0}) and thus of
$\bar{\tau}_1$, (\ref{33.1}) can be related to the $\tau$-function
sequence (\ref{4.2}).

\begin{prop}\label{p.6}
Of the two linearly independent solutions to (\ref{s2.0}), choose the solution
analytic at the origin,
\begin{equation}\label{44.m}
e^{-(v_1^{(0)} - v_3^{(0)})t} \tau_1^{V^*} = {}_1 F_1 (v_1^{(0)} - v_3^{(0)},
v_1^{(0)} - v_2^{(0)}+1;-t)
\end{equation}
with
\begin{equation}\label{44.m1}
v_1^{(0)}- v_3^{(0)} = - \nu, \qquad v_1^{(0)}- v_2^{(0)} = \mu.
\end{equation}
Then (\ref{33.0}) and (\ref{33.1}) give
\begin{equation}\label{3.1.a}
e^{\nu t} \tau_n^{V^*} = \Big ( {\Gamma(\mu+1) \over \Gamma(\mu+\nu+1)}
\Big )^n \prod_{l=0}^{n-1} \Gamma(\nu+l+1) \,
\tau^V[n](t;\mu,\nu).
\end{equation}
\end{prop}

\noindent
Proof. \quad Choosing $\tau_1^{V^*}$ as in (\ref{44.m}) and the parameters as
in (\ref{44.m1}) gives, upon comparing with (\ref{T5.1})
$$
e^{\nu t} \tau_1^{V^*} = {\Gamma(\mu+1) \Gamma(\nu +1) \over
\Gamma(\mu+\nu+1)} \tau^V[1](t;\mu,\nu).
$$
Furthermore, use of (\ref{33.0}) and (\ref{33.1}) shows
\begin{eqnarray}
\lefteqn{
t^{n^2/2} e^{(\nu + n) t} \tau_n^{V^*}} \nonumber \\&&
= \Big ( {\Gamma(\mu+1) \Gamma(\nu+1) \over 2 \pi \Gamma(\mu+\nu+1)}
\Big )^n
\det \Big [ \delta^{j+k} t^{1/2} e^t
\int_{-\pi}^\pi z^{(\mu-\nu)/2}|1+z|^{\mu+\nu}e^{tz} \, d\theta
\Big ]_{j,k=0,\dots,n-1}.
\label{44.0}
\end{eqnarray}
We know \cite[proof of Prop.~3.1]{FW01a} that (\ref{33.1}) is equivalent to
\begin{equation}\label{44.1}
t^{nc} \bar{\tau}_n = \det [ \delta^{j+k} t^c \bar{\tau}_1]_{j,k=0,\dots,n-1}
\end{equation}
for any $c$, and thus choosing $c=-1/2$ the factor of $t^{1/2}$ in the
determinant is cancelled, while the left hand side of (\ref{44.0}) is
multiplied by $t^{-n/2}$. Now simple manipulation involving
integration by parts shows
\begin{eqnarray}\label{44.2}
\lefteqn{\delta \Big ( e^t \int_{-\pi}^\pi(1+z)^\mu(1+1/z)^\nu e^{tz}
\, d\theta \Big )} \nonumber \\
&&
= e^t \Big ( (\nu+1) \int_{-\pi}^\pi (1+z)^\mu (1+1/z)^{\nu+1}
e^{tz} \, d\theta - (\mu+\nu+1)
\int_{-\pi}^\pi(1+z)^\mu(1+1/z)^\nu e^{tz}
\, d\theta \Big ),
\end{eqnarray}
so making use of (\ref{4.3e}) and applying (\ref{44.2}) to column $k$ and
subtracting $(\mu+\nu+1)$ times column $k-1$ for $k=n-1,n-2,\dots,1$ in
order shows (\ref{44.0}) can be reduced to
\begin{eqnarray*}
\lefteqn{
t^{n(n-1)/2} e^{(\nu + n) t} \tau_n^{V^*} 
= \Big ( {\Gamma(\mu+1) \Gamma(\nu+1) \over 2 \pi \Gamma(\mu+\nu+1)}
\Big )^n (\nu+1)^{n-1}} \nonumber \\&&
\times
\det \Big [ \delta^{j}  e^t
\int_{-\pi}^\pi (1+z)^\mu(1+1/z)^\nu e^{tz} \, d\theta
\quad
\delta^{j+k-1} e^t \int_{-\pi}^\pi (1+z)^\mu(1+1/z)^{\nu+1} e^{tz} \, d\theta
\Big ]_{j=0,\dots,n-1 \atop k=1,\dots,n-1}.
\end{eqnarray*}
Repeating this procedure for columns $k=n-1,n-2,\dots,2$ and so on shows
\begin{eqnarray*}
\lefteqn{
t^{n(n-1)/2} e^{(\nu + n) t} \tau_n^{V^*}} \nonumber \\&&
 =
\Big ( {\Gamma(\mu+1)  \over 2 \pi \Gamma(\mu+\nu+1)}
\Big )^n \prod_{l=0}^{n-1}\Gamma(\nu+l+1)
\det \Big [ \delta^{j}  e^t
\int_{-\pi}^\pi (1+z)^\mu(1+1/z)^{\nu+k} e^{tz} \, d\theta
\Big ]_{j,k=0,\dots,n-1},
\end{eqnarray*}
and application of the general identities \cite{FW01a}
\begin{eqnarray*}
\det \Big [ \delta^j(u(t)f_k(t)) \Big ]_{j,k=0,\dots,n-1} & = &
(u(t))^n \det [ \delta^j f_k(t) ]_{j,k=0,\dots,n-1} \\
\det \Big [ \delta^j f_k(t) \Big ]_{j,k=0,\dots,n-1} & = &
t^{n(n-1)/2} \det \Big [ {d^j \over dt^j} f_k(t) \Big ]_{j,k=0,\dots,n-1}
\end{eqnarray*}
then gives
\begin{equation}\label{4.18m}
 e^{\nu  t} \tau_n^{V^*} =
\Big ( {\Gamma(\mu+1)  \over 2 \pi \Gamma(\mu+\nu+1)}
\Big )^n \prod_{l=0}^{n-1}\Gamma(\nu+l+1)
\det \Big [ 
\int_{-\pi}^\pi (1+z)^\mu(1+1/z)^{\nu+k} z^j e^{tz} \, d\theta
\Big ]_{j,k=0,\dots,n-1}.
\end{equation}
The stated result now follows after noting that the factor
$(1+1/z)^{k}$ in the integral can be replaced by
$(1/z)^{k}$ without changing the value of the determinant.
\hfill $\square$

\medskip
Knowledge of (\ref{3.1.a}) and key recurrences from the Okamoto theory
of PV as detailed in \cite{FW01a} allows the following recurrence for
$\tau^V[N](t;\mu,\nu)$ to be deduced.

\begin{prop}\label{p.7}
Let $\tau^V[N]= \tau^V[N](t;\mu,\nu)$ as given by (\ref{4.2})
or more generally (\ref{T5.1}). Let $p_N, q_N$ denote the conjugate
variables in the Hamiltonian (\ref{s1.1}) with parameters
\begin{equation}\label{66.1}
v_1 - v_3 = -\nu, \quad v_1-v_2=\mu, \quad v_4-v_3=N
\end{equation}
and define
\begin{equation}\label{66.1a}
x_N := (p_N +t) q_N - {1 \over 2} (v_2 - v_1), \qquad y_N = {1 \over q_N}.
\end{equation}
The sequences $\{\tau^V[N]\}_{N=0,1,\dots}$, $\{x_N\}_{N=0,1,\dots}$,
$\{y_N\}_{N=0,1,\dots}$ satisfy the coupled recurrences
\begin{eqnarray}
(N+\nu)  {\tau^V[N+1] \tau^V[N-1] \over (\tau^V[N])^2 } & = &
\Big ( x_N - {t \over y_N} - \nu - {\mu \over 2} \Big )
\Big ( {1 \over y_N} - 1 \Big ) + N \\
x_N + x_{N-1} & = & {t \over y_N} - {N \over 1 - y_N} \label{66.3} \\
y_N y_{N+1} & = & t {x_N + \nu + \mu/2 + N +1 \over x_N^2 - (\mu/2)^2}
\label{66.4}
\end{eqnarray}
subject to the initial conditions
\begin{eqnarray}
&& x_0 = t + \mu/2 + t {d \over dt} \log {}_1F_1(-\nu,\mu+1;-t), \qquad
y_0 =1,  \label{v.1}
\\
&& \tau^V[0] =1, \qquad
\tau^V[1] = {\Gamma(\mu+\nu+1) \over \Gamma(\mu+1) \Gamma(\nu+1) }
{}_1F_1(-\nu,\mu+1;-t). \label{v.2}
\end{eqnarray}
\end{prop}

\noindent
Proof. \quad According to \cite[proof of Prop.~2.2]{FW01a}
\begin{equation}\label{4.24a}
{\tau_{n+1}^{V^*} \tau_{n-1}^{V^*} \over (\tau_{n}^{V^*})^2} =
{\partial \over \partial t} \Big ( t H^{V^*} + (v_4^{(0)} - v_1^{(0)} + n)t
\Big )
\end{equation}
so taking into consideration (\ref{3.1.a}), (\ref{s1.1}) and (\ref{66.1})
we arrive at (\ref{66.3}), (\ref{66.4}). 
Furthermore in \cite[Prop.~2.4]{FW01a} it was shown
that with $x_N, y_N$ specified in terms of $p_N$, $q_N$ by (\ref{66.1a}),
$\{x_N, y_N\}$ satisfy the discrete Painlev\'e IV recurrences
\begin{eqnarray}\label{4.26a}
x_N + x_{N-1} & = & {t \over y_N} + 
{v_3^{(0)} - v_4^{(0)} \over 1  - y_N} \nonumber 
\\
y_N y_{N+1}  & = & t 
{x_N - {1 \over 2}(v_1^{(0)}+v_2^{(0)}) + 1 + v_4^{(0)} + N \over
x_N^2 - {1 \over 4} (v_2^{(0)} - v_1^{(0)})^2 }.
\end{eqnarray}
Making use of  (\ref{66.1}) then gives (\ref{66.3}) and (\ref{66.4}).
The initial conditions follow from (\ref{66.1a}), (\ref{s1.5}) and
(\ref{44.m}).
\hfill $\square$

For $n \in \zz_{\ge 0}$, the confluent hypergeometric function
${}_1 F_1(-n;c;-t)$ is proportional to a Laguerre polynomial,
$$
{}_1 F_1(-n;\alpha+1;-t) = {\Gamma(\alpha+1) \Gamma(n+1) \over
\Gamma(n+\alpha+1)} L_n^\alpha(-t).
$$
Thus for $\nu \in \zz_{\ge 0}$ it follows from (\ref{T5.1}) that
\begin{equation}\label{ni.1}
\tau^V[N](t;\mu,\nu) = \det \Big [ L_{\nu+k-j}^{\mu+j-k}(-t)
\Big ]_{j,k=1,\dots,N}
\end{equation}
(note that $L_n^\alpha(-t) := 0$ for $n < 0$). According to
(\ref{v.1}), (\ref{v.2}) we also have
\begin{equation}\label{ni.2}
x_0 = t + \nu + \mu/2 - (\nu + \mu) {L_{\nu-1}^\mu(-t) \over
L_\nu^\mu(-t)}, \quad y_0=1, \quad \tau^V[0]=1, \quad
\tau^V[1] = L_\nu^\mu(-t).
\end{equation}

As in the PIV theory, in the case $\nu \in \zz_{\ge 0}$ the $N \times
N$ determinant for $\tau^V[N]$ can also be expressed as a 
$\nu \times \nu$ determinant. Thus we know from
\cite[Props.~3.6,3.7]{FW01a} that for $\nu \in \zz_{\ge 0}$
\begin{equation}\label{ni.3}
\tau^V[N](t;\mu,\nu) \propto \det \Big [ {d^j \over d t^j}
L_{N+k}^\mu(-t) \Big ]_{j,k=0,\dots,\nu-1}.
\end{equation}
Using the Laguerre polynomial identities
$$
L_n^{\alpha-1}(x) = L_{n}^{\alpha}(x) - L_{n-1}^{\alpha}(x), \qquad
{d \over dx} L_p^{\alpha}(x) = - L_{p-1}^{\alpha+1}(x)
$$
this is equivalent to
$$
\tau^V[N](t;\mu,\nu) \propto \det  \Big [ L_{N+k-j}^{\mu+j-k}(-t)
\Big ]_{j,k=1,\dots,\nu}
$$
and thus
$$
\tau^V[N](t;\mu,\nu) \propto \tau^V[\nu](t;\mu,N).
$$
To determine the proportionality, we use the fact, following from
(\ref{4.2}) and (\ref{4.3e}), that
$$
\tau^V[N](0,\mu,\nu) = {1 \over N!} M_N(\mu,\nu) 
$$
where
\begin{eqnarray}\label{m.n}
M_N(a,b) & := & 
\int_{-1/2}^{1/2} dx_1 \cdots \int_{-1/2}^{1/2} dx_N \,
\prod_{l=1}^N u_l^{(a-b)/2}|1+u_l|^{a+b} \prod_{1 \le j < k \le N}
|u_k - u_j|^2, \quad u_l := e^{2 \pi i x_l}. \nonumber \\
& = & \prod_{j=0}^{N-1} {\Gamma(a+b+1+j) \Gamma(2+j) \over
\Gamma(a+1+j) \Gamma(b+1+j)}.
\end{eqnarray}
Hence for $\nu \in \zz_{\ge 0}$
\begin{equation}\label{4.31}
\tau^V[N](t;\mu,\nu) = {\nu! \over N!} {M_N(\mu,\nu) \over
M_\nu(\mu,N)} \tau^V[\nu](t;\mu,N).
\end{equation}

Another point of interest is the explicit $\tau$-function form of the
sequences $\{x_N\}_{N=0,1,\dots}$, $\{y_N\}_{N=0,1,\dots}$ generated
by the discrete Painlev\'e IV coupled recurrences (\ref{66.3}),
(\ref{66.4}) with initial conditions (\ref{v.1}). We find
\begin{eqnarray}\label{4.33}
x_N - \mu/2 & = & t {\tau[N+1](t;\mu+1,\nu) \tau[N](t;\mu-1,\nu+1)
\over \tau[N+1](t;\mu, \nu) \tau[N](t;\mu, \nu+1) } \nonumber \\
y_N & = & {\tau[N](t;\mu,\nu+1) \tau[N](t;\mu,\nu) \over
\tau[N](t;\mu-1,\nu+1) \tau[N](t;\mu+1,\nu)}.
\end{eqnarray}
These formulas can be established by proceeding in an analogous fashion
to the proof of Proposition \ref{pr.4}. 

Let us now turn our attention to the PV $\tau$-function
$\tilde{\tau}^V$ as specified by (\ref{1.7}). In some special
cases this is intimately related to $\tau^V$ as specified by
(\ref{4.2a}). Thus with $I_N(a)$ specified by (\ref{I.N})
we have \cite[Prop.~3.7]{FW01a} 
\begin{eqnarray}
{I_N(a) \over I_N(a+\mu)} \tilde{\tau}^V[N](-t;\mu,a;0) & = &
{M_\mu(0,0) \over M_\mu(a,N)}
{\tau}^V[\mu](t;a,N), \qquad \mu \in \zz_{\ge 0} \label{I.1}\\
{I_N(a) \over I_N(a+\mu)} \tilde{\tau}^V[N](t;\mu,a;1) & = &
{M_a(0,0) \over M_a(\mu,N)}
\tau^V[a](t;\mu,N), \qquad a \in \zz_{\ge 0} \label{I.2}
\end{eqnarray}

The identities (\ref{I.1}) and (\ref{I.2}) allow those special cases
of $\tilde{\tau}^V[n]$ to be computed by the recurrences of 
Proposition \ref{p.7}, however the recurrences will no longer
be with respect to
the dimension of the average $n$ but rather with respect to one
of the parameters. For general parameters a system of
recurrences for $\tilde{\tau}^V[n]$ can also be given, but these
recurrences alter both $n$ and the parameter $a$. This comes
about as a consequence of the following analogue of
Proposition \ref{p.6}.

\begin{prop}
Write the general solution of the confluent hypergeometric equation
(\ref{s2.0}) in the integral form
$$
e^{-(v_1^{(0)} - v_3^{(0)})t} \tau_1^{V^*} =
\Big ( \int_0^\infty - \xi \int_0^1 \Big ) e^{-tu}
(u-1)^{v_2^{(0)} - v_3^{(0)}} u^{v_1^{(0)} - v_2^{(0)}-1} \, du,
$$
and set
$$
\alpha =: v_2^{(0)} - v_3^{(0)}+1, \qquad
\gamma =: v_1^{(0)} - v_3^{(0)}+1.
$$
We have
\begin{equation}\label{ni.7}
\tau_n^{V^*} = e^{(\gamma-1)t} t^{- n(\gamma-1)}
(\Gamma(\gamma-\alpha))^{n-1} \prod_{l=0}^{n-1} \Gamma(l+1)
\tilde{\tau}^V[n](t;\alpha -1, \gamma - \alpha - n;\xi).
\end{equation}
\end{prop}

\noindent
\quad Proof. With
$$
F(\alpha,\gamma;t) := e^t 
\Big ( \int_0^\infty - \xi \int_0^1 \Big ) e^{-tu}
(u-1)^{\alpha - 1} u^{\gamma - \alpha -1} \, du
$$
we see from (\ref{33.0}), (\ref{33.1}) and (\ref{44.1}) that
$$
t^{n(n-1)/2} e^{(-\gamma +1 + n)t} \tau_n^{V^*} =
\det [ \delta^{j+k} F(\alpha,\gamma;t) ]_{j,k=0,\dots,n-1}.
$$
Analogous to (\ref{44.2}) we can show that
$$
\delta F(\alpha,\gamma;t) = - \alpha F(\alpha,\gamma;t) -
(\gamma - \alpha - 1) F(\alpha+1,\gamma;t).
$$
Then proceeding as in the derivation of (\ref{4.18m}) we deduce
$$
e^{-(\gamma-1)t} \tau_n^{V^*} = 
{(\Gamma(\gamma-\alpha))^{n-1} \over
\prod_{l=1}^{n-1} \Gamma(\gamma-\alpha-l)}
\det [ e^{-t} F(\alpha + k, \gamma+j;t) ]_{j,k=0,\dots,n-1}.
$$
The method of the proof of \cite[proof of Prop.~3.1]{FW01a} allows this 
determinant to be written as a multiple integral. Making use too
of (\ref{I.N}) then gives (\ref{ni.7}). \hfill  $\square$ 

\smallskip
Because (\ref{4.24a}) and (\ref{4.26a}) hold for any $\tau$-function
sequence with the property (\ref{s1.2}), we see that (\ref{ni.7})
allows us to specify the analogue of Proposition \ref{p.7} for
$\{ \tilde{\tau}^V[n](t;\alpha-1,\gamma-\alpha-n;\xi) \}_{n=0,1,\dots}$,
although we stop short of writing it down.

\section{The $\tau$-function sequence $\{\tau^{VI}[n]
(t;\mu,w_1,w_2;\xi)\}_{n=0,1,
\dots}$}
\setcounter{equation}{0}
As written (\ref{4.3}) requires $-\pi < \phi \le \pi$ to make sense, however
we can readily extend this definition to general complex
\begin{equation}\label{m.0}
t := e^{i \phi}
\end{equation}
First we make use of (\ref{11.0}) to obtain
\begin{equation}\label{m.1}
\tau^{VI}[N] = \det \Big [ {1 \over 2 \pi} \int_{-\pi}^\pi
(1 - \xi \chi_{(\pi - \phi, \pi)}) e^{w_2 \theta} |1 + z|^{2 w_1}
| e^{i(\pi - \phi)} - z|^{2 \mu} z^{j-k} \, d \theta
\Big ]_{j,k=1,\dots,N}.
\end{equation}
The integral in (\ref{m.1}) naturally breaks into two. Introducing $t$
according to (\ref{m.0}) and setting $\mu \in \zz_{\ge 0}$ the first portion
reads
\begin{eqnarray}
&& {1 \over 2 \pi} \int_{-\pi}^\pi e^{w_2 \theta} 
|1 + z|^{2 w_1} \Big ( {1 \over t z} \Big )^\mu (1 +tz)^{2 \mu}
z^{j-k} \, d \theta 
= {\Gamma(a+b+1) \over \Gamma(a+1) \Gamma(b+1)} t^{-\mu}
\, {}_2 F_1(-2 \mu,-b;a+1;t) \label{m.1.a}
\end{eqnarray}
where, with $w := w_1 + i w_2$,
$$
a = \bar{w} - \mu + j - k, \qquad b=w + \mu -j+k
$$
(the integral evaluation follows upon making use of (\ref{fbo})). For the
second portion, taking both $w_1, \mu \in \zz_{\ge 0}$, and writing in
terms of $dz$ instead of $d \theta$ we have
\begin{equation}\label{m.2.1}
- {\xi t^{-\mu} \over 2 \pi i} \int_{{\cal C}_{(-1/t,-1)}}
z^{-i w_2 + j - k - w_1 - \mu} (1 + z)^{2 w_1} (1 + tz)^{2 \mu} \,
{dz \over z}
\end{equation}
where ${\cal C}_{(-1/t,-1)}$ is a simple closed contour starting at
$z=-1/t$ and finishing at $z=-1$. Making the successive transformations
$z \mapsto -z/t$, $z \mapsto -z+1$, $z \mapsto (1-t)z$, then making use
of the integral formula
$$
\int_0^1 x^{\lambda_1}(1-x)^{\lambda_2} (1 - tx)^{-r} \, dx =
{\Gamma(\lambda_1+1) \Gamma(\lambda_2+1) \over \Gamma(\lambda_1+
\lambda_2+2)} \, {}_2F_1(r,\lambda_1+1,\lambda_1+\lambda_2+2;t)
$$
shows (\ref{m.2.1}) is equal to
\begin{eqnarray}
&&{\xi t^{-\mu}\over 2 \pi i} e^{\pm \pi i (k-j+\mu- \bar{w})}
{\Gamma(2 \mu +1) \Gamma(2w_1+1) \over \Gamma(2 \mu+2w_1+2)}
t^{k-j+\mu- \bar{w}} (1 - t)^{2\mu+2w_1+1}  \nonumber \\
&& \quad \times {}_2 F_1(2\mu+1,1+k-j+\mu+w;2\mu+2w_1+2;1-t)
\label{m.2.a}
\end{eqnarray}
where the $\pm$ sign is taken accordingly as
Im$(t) \lessgtr 0$. Substituting for the integral in (\ref{m.1}) the
sum of the hypergeometric functions (\ref{m.1.a}) and (\ref{m.2.a})
gives meaning to $\tau^{VI}[N]$ for general complex values of
(\ref{m.0}).

We know from \cite{FW01a} that the CUE${}_N$ average (\ref{4.3}) can be
written as an average over the generalized Cauchy unitary ensemble
\cite{WF00a,BO01} specified by the p.d.f. 
\begin{equation}\label{5.48}
{1 \over C}\prod_{l=1}^N {1 \over (1+i x_l)^\eta (1-i x_l)^{\bar{\eta}}}
\prod_{1 \le j < k \le N} (x_k - x_j)^2, \qquad
C = 2^{-N(N-1)} \pi^N M_N(\bar{\eta} - N, \eta - N)
\end{equation}
where $M_N$ is given by (\ref{m.n}).
Thus making the change of variables
$$
z_l = {1 + ix_l \over 1 - ix_l }
$$
in (\ref{4.3}) shows \cite[eq.~(1.19) with $\mu \mapsto 2\mu$]{FW02}
\begin{equation}\label{fc.1}
{  N! \over M_N(\mu+\bar{w},\mu+w)} 
\tau^{VI}[N](e^{i \phi} ;\mu,w_1,w_2;\xi^*) 
\quad = {1 \over (1 + s^2)^{N \mu}}
\Big \langle \prod_{l=1}^N(1 - \xi \chi_{(s,\infty)}^{(l)} )
(s - x_l)^{2 \mu} \Big \rangle_{{\rm CyUE}_N}
\Big |_{\eta = w+\mu +N, \atop
s = \cot \phi/2}
\end{equation}
where
\begin{equation}
\xi^*  :=  1 - (1-\xi)e^{-\pi i \mu}
\end{equation} 
Furthermore, we know from \cite{FW02} that the CyUE average in (\ref{fc.1})
as a function of $s$ is also the $\tau$-function for a particular
PVI system. Moreover, unlike the situation with (\ref{4.3}), the Schlesinger
transformation studied in \cite{FW02} increments $N$ in this average
and leaves the other parameters unchanged (in (\ref{4.3}) this same
Schlesinger transformation increments $N$ but also decrements $\mu$).

To make these statements more explicit, we recall \cite{Ok87b} that the
Hamiltonian for the PVI system is given by
\begin{eqnarray}\label{5.49}
&&t(t-1)H^{VI} = q(q-1)(q-t)p^2 - \Big ( (v_1+v_2)(q-1)(q-t) 
\nonumber \\
&& \quad + (v_1-v_2)q(q-t) + (v_3+v_4) q(q-1) \Big ) p +
(v_1+v_4)(v_1+v_3)(q-t).
\end{eqnarray} 
Introduce the Schlesinger operator $T_3$ with action on the parameters
$$
T_3 \mathbf v = (v_1, v_2, v_3+1, v_4),
$$
and with appropriate actions on the conjugate variables $p,q$. We know
\cite[eq.~(2.27)]{FW02} that
$$
T_3^n H_0^{VI} =: H_n^{VI} = H_0^{VI} \Big |_{\mathbf v \mapsto
(v_1^{(0)}, v_2^{(0)}, v_3^{(0)}+n, v_4^{(0)})}.
$$
>From this we introduce a sequence of $\tau$-functions $\tau_n^{VI}$ specified
so that
$$
H_n^{VI} = {d \over dt} \log \tau_n^{VI}.
$$
We know from \cite[Prop.~15]{FW02} that
\begin{equation}\label{m.3b1}
\Big \langle  \prod_{l=1}^N(1 - \xi \chi_{(s,\infty)}^{(l)})
(s - x_l)^\mu \Big \rangle_{{\rm CyUE}_N} \Big |_{\eta \mapsto N+\eta}
\propto
\tau_N^{VI} \Big ( {is+1 \over 2}; \mathbf v\Big )
\end{equation}
where with $\eta := \eta_1 + i \eta_2$
\begin{equation}\label{m.3b2}
\mathbf v = (-\eta_1, i\eta_2, \eta_1+N, - \mu+\eta_1).
\end{equation}
For our present purposes the proportionality constant in (\ref{m.3b1}), not
calculated in \cite{FW02}, is of importance. To obtain its value we must
recall some of the results from \cite{FW02}.

We know from \cite[eqs.~(2.30),(2.37)]{FW02} that with
\begin{equation}\label{m.3c.1}
\bar{\tau}_n^{VI} := (t(t-1))^{(n+v_1^{(0)}+v_3^{(0)})(n+v_3^{(0)}+
v_4^{(0)})/2} \tau_n^{VI}
\end{equation}
we have
\begin{equation}\label{m.3c.2}
\bar{\tau}_n^{VI} = \det [ \delta^{j+k} \bar{\tau}_1^{VI} ]_{j,k=0,\dots,
n-1}, \qquad \delta := t(t-1) {d \over dt}.
\end{equation}
We also know that $\tau_1^{VI}$ satisfies the Gauss hypergeometric
differential equation
\begin{equation}\label{m.3c.3}
t(1-t) y'' + \Big (c - (a+b+1)t \Big ) y' - ab y = 0
\end{equation}
with
\begin{equation}\label{m.3c.4}
a = v_4^{(0)} - v_3^{(0)}, \quad b=1+ v_3^{(0)} + v_4^{(0)}, \quad
c = 1 + v_2^{(0)} + v_4^{(0)}
\end{equation}
and that a general solution of (\ref{m.3c.3}) is given by
 \cite[eqs.~(2.67)]{FW02}
\begin{equation}\label{m.3c.5}
\tau_1^{VI} = F(a,b,c;t) = \Big ( \int_{-\infty}^\infty - \xi
\int_t^\infty \Big ) u^{a-c}(1 - u)^{c-b-1} (t-u)^{-a} \, du.
\end{equation}
Noting that in (\ref{m.3b2}) we have $ v_1^{(0)} + v_3^{(0)} = 0$, it
thus follows from (\ref{m.3c.1}), (\ref{m.3c.2}), (\ref{m.3c.4}) and
(\ref{m.3c.5}) that
\begin{equation}\label{m.3c.6}
\bar{\tau}_n^{VI} = \det [ \delta^{j+k}(t(t-1))^{b/2} F(a,b,c;t) ]_{j,k=0,
\dots, n-1}.
\end{equation}

We know from \cite[proof of Prop.~6]{FW02} that
$F$ satisfies the differential-difference
relations
\begin{eqnarray}
t {d \over dt} F(a,b,c;t) & = & -(c-b-1) F(a,b+1,c;t) - b F(a,b,c;t)
\label{m.3c.7a} \\
t(1-t) {d \over dt} F(a,b,c;t) & = & (a-c) F(a-1,b,c;t) +
(a-c+bt) F(a,b,c;t). \label{m.3c.7b}
\end{eqnarray}
It follows that
\begin{eqnarray*}
\delta((t(t-1))^{b/2} F(a,b,c;t) ) & = & {b \over 2}
(t(t-1))^{b/2} F(a,b,c;t) + (b+1-c)t^{b/2}(t-1)^{b/2+1} F(a,b+1,c;t) \\
\delta((t(t-1))^{b/2} F(a,b,c;t) ) & = &
(c-a-b/2)(t(t-1))^{b/2} F(a,b,c;t) + (c-a)(t(t-1))^{b/2} 
F(a-1,b,c;t).
\end{eqnarray*}
>From these latter relations
the working of the proof of \cite[Prop.~4]{FW02} gives
\begin{equation}
 \bar{\tau}_n^{VI}  =   \prod_{j=1}^{n-1}(b+1-c)_j
(c-a)_j 
t^{bn/2}(t-1)^{bn/2+n(n-1)/2} \det \Big [ F(a-j,b+k,c;t)
\Big ]_{j,k=0,\dots,n-1}.
\end{equation}
The method of the proof of \cite[Prop.~5]{FW02} allows this to be rewritten
as a multiple integral, which when substituted in (\ref{m.3c.1}) and
after substitution of the parameters according to (\ref{m.3c.4}),
(\ref{m.3b2}) shows
\begin{eqnarray}\label{5.60}
\tau_n^{VI} & = & {(-1)^{n(n-1)/2} \over n!} \prod_{j=1}^{n-1}
(1 + \bar{\eta})_j (1 + \eta)_j \Big ( \int_{-\infty}^\infty - \xi
\int_{t}^\infty \Big ) du_1 \cdots \Big ( \int_{-\infty}^\infty - \xi
\int_{t}^\infty \Big ) du_n \nonumber \\
&& \times \prod_{j=1}^n u_j^{-(\eta + n)}(1 - u_j)^{-(\bar{\eta} + n)}
(t - u_j)^\mu \prod_{1 \le j < k \le n} (u_k - u_j)^2.
\end{eqnarray}
Replacing $t$ by $(is+1)/2$, changing variables in the integrations
$u_j \mapsto (iv_j+1)/2$ and making use of (\ref{5.48})
we deduce that the  proportionality constant in 
(\ref{m.3b1}) (taken for convenience to be on the left hand side) is
equal to
\begin{equation}\label{5.61}
\Big ( {1 \over N!} \prod_{j=1}^{N-1} (1 + \bar{\eta})_j(1+\eta)_j
\Big ) i^{(\mu+1)N} 2^{-(\mu-2 \eta_1 -1)N} 
\pi^n M_N(\bar{\eta},\eta).
\end{equation}
Consequently, after recalling (\ref{fc.1}), we conclude
\begin{eqnarray}\label{5.62}
&& \Big ( {1 \over (1 + s^2)^{N \mu/2}}
\tau_N^{VI} \Big ( {is+1 \over 2}; \mathbf{v} \Big ) \Big ) \Big |_{s = \cot
\phi/2, \, \mu \mapsto 2 \mu \atop \eta = w+ \mu} \nonumber 
\\
&& \quad = \Big ( \prod_{j=1}^{N-1} (1  + \bar{w}+\mu)_j(1+w+\mu)_j
\Big )  i^{(\mu+1)N} 2^{-(2w_1+1)N} \pi^N 
\tau^{VI}[N](e^{i \phi};\mu,w_1,w_2;\xi^*).
\end{eqnarray}

The use of this result lies with the fact that the $\tau$-function
sequence $\{\tau_N^{VI}(t;\mathbf{v})\}_{N=0,1,\dots}$ for general
$t$ and $\mathbf{v}=(v_1^{(0)},v_2^{(0)},v_3^{(0)}+N,v_4^{(0)})$
satisfies, according to \cite[proof of Prop.~2]{FW02}, an
equation of the form (\ref{5.3}),
\begin{eqnarray}\label{5.23a}
&&{\tau_{N+1}^{VI} \tau_{N-1}^{VI} \over (\tau_N^{VI})^2 }  = 
{\partial \over \partial t} \Big ( t(t-1) H^{VI} \Big ) +
(v_1^{(0)} + v_3^{(0)} +N)( v_3^{(0)} +  v_4^{(0)} + N) \nonumber
\\ && \quad = q_N(1-q_N)p_N^2 + 2 v_1^{(0)} q_N p_N - (v_1^{(0)} +
v_2^{(0)}) p_N + (v_1^{(0)} +  v_3^{(0)} + N)(v_3^{(0)} - v_1^{(0)} + N),
\end{eqnarray}
and in addition recurrences determining $\{p_n\}_{n=1,2,\dots}$ and
$\{q_n\}_{n=1,2,\dots}$ are also known. Regarding the latter, set
\begin{eqnarray}\label{5.63}
g_n & := & {q_n \over q_n - 1}, \nonumber \\
f_n & := & q_n(q_n-1) p_n + (1+n-\alpha_2^{(0)}-\alpha_4^{(0)})(q_n-1)
- \alpha_3^{(0)} q_n - (\alpha_0^{(0)} + n) {q_n(q_n-1) \over q_n - t}
\end{eqnarray}
where
\begin{equation}\label{5.64}
\alpha_0^{(0)} = v_3^{(0)}+v_4^{(0)}+1, \: \:
\alpha_1^{(0)} = v_3^{(0)}-v_4^{(0)}, \: \:
\alpha_2^{(0)} = -(v_1^{(0)}+v_3^{(0)}), \: \:
\alpha_3^{(0)} = v_1^{(0)}-v_2^{(0)}, \: \:
\alpha_4^{(0)} = v_1^{(0)}+v_2^{(0)}. 
\end{equation}
Then we have \cite[Prop.~10]{FW02}
\begin{eqnarray}\label{5.65}
g_{n+1} g_n & = & {t \over t - 1} {(f_n+1 +n - \alpha_2^{(0)})
(f_n +1 +n - \alpha_2^{(0)} - \alpha_4^{(0)}) \over f_n(f_n + \alpha_3^{(0)}) }
\nonumber \\
f_n + f_{n-1}  & = & - \alpha_3^{(0)} + {\alpha_1^{(0)}+n  \over g_n - 1}
+ {(\alpha_0^{(0)}+n) t \over t(g_n - 1) - g_n},
\end{eqnarray}
which are a version of the discrete Painlev\'e V equations \cite{Sa01}.
To use these recurrences to determine $\{p_n\}$, $\{q_n\}$ given $f_0, g_0$
we first iterate (\ref{5.65}) to determine $\{f_n\}, \{g_n\}$.
According to the first equation in (\ref{5.63}) $q_n$ can then be
calculated in terms of $g_n$,
\begin{equation}\label{5.66}
q_n = {g_n \over g_n - 1}.
\end{equation}
Now that $q_n$ is known the second equation in (\ref{5.63}) allows $p_n$
to be calculated in terms of $f_n, q_n$,
\begin{eqnarray}\label{5.67}
&& p_n = {1 \over q_n(q_n-1)} \Big (
f_n + (1+n-\alpha_2^{(0)} - \alpha_4^{(0)})(1-q_n) + \alpha_3^{(0)}
q_n   
 + (\alpha_0^{(0)} + n) {q_n(q_n-1) \over q_n - t} \Big ).
\end{eqnarray}
To calculate $\{\tau^{VI}[N](e^{i \phi};\mu,w_1,w_2;\xi) \}_{N=2,3,\dots}$
the following recurrence scheme can therefore be given.

\begin{prop}\label{p.8}
Let $\tau^{VI}[N]:= \tau^{VI}[N](e^{i \phi};\mu,w_1,w_2;\xi)$ as specified
by (\ref{m.1}) and let $p_N$, $q_N$ denote the conjugate variables in the
Hamiltonian (\ref{5.49}) with parameters (\ref{m.3b2}). We have
\begin{eqnarray}\label{5.29}
q_n & = & {g_n \over g_n - 1} \nonumber \\
p_n & = & {1 \over q_n(q_n-1)} \Big ( f_n + (1+n+\bar{w}+\mu)(1 - q_n)
- (w+\mu) q_n  \nonumber \\
&&
+ (1 + n + 2w_1) {q_n(q_n-1) \over q_n - (1-e^{i \phi})^{-1}}
\Big )
\end{eqnarray}
where $\{f_n\}_{n=0,1,\dots}$, $\{g_n\}_{n=0,1,\dots}$ are determined by
the recurrences
\begin{eqnarray}
g_{n+1} g_n & = & e^{-i \phi} 
{(f_n+1+n) (f_n+1 +n + \bar{w}+\mu) \over f_n ( f_n - w-\mu))} \nonumber \\
f_n+f_{n-1} & = & {w}+\mu + {2 \mu +n \over g_n - 1} -
{(1+n+2 w_1) \over 1 - e^{i \phi} g_n}
\end{eqnarray}
subject to the initial conditions
\begin{equation}\label{5.in}
g_0 = {q_0 \over q_0 - 1}, \qquad
f_0 = (1 + \bar{w} + \mu)(q_0 - 1) + (w + \mu) q_0 - (2w_1 + 1)
{q_0 (q_0 - 1) \over q_0 - (1 - e^{i \phi})^{-1}}
\end{equation}
with
\begin{equation}\label{5.32}
q_0 = {1 \over 2}\Big ( 1  + {i \over  \mu} {d \over d \phi} \log \tau^{VI}[1]
\Big ).
\end{equation}
Given $\tau^{VI}[0]=1$, and $\tau^{VI}[1]$ as the element of the determinant
in   (\ref{m.1}) with $j-k=0$, we have that
$\{\tau^{VI}[N]\}_{N=2,3,\dots}$ can be computed in terms of
$\{p_N\}_{N=1,2,\dots}$ and $\{q_N\}_{N=1,2,\dots}$ by the recurrence
\begin{eqnarray}\label{5.34}
&&  (N+ \bar{w} + \mu)(N+w+\mu) 
{ \tau^{VI}[N+1] \tau^{VI}[N-1] \over (\tau^{VI}[N])^2}
\nonumber \\
&& \quad =
q_N(1-q_N) p_N^2 - 2(w_1 + \mu) q_N p_N + (\bar{w} + \mu)p_N +
N(N+2w_1+2 \mu).
\end{eqnarray}
\end{prop}

\noindent
Proof. \quad The only remaining point to require explanation is the initial
conditions (\ref{5.in}), (\ref{5.32}). These come about because the PVI
system admits the solution \cite[Prop.~3]{FW02} 
\begin{equation}\label{5.33a}
p_0=0, \qquad t(t-1) {d \over dt} \log \tau_1^{VI}(t) = - \mu (q_0-t).
\end{equation}
According to (\ref{5.62}) we require $t = {1 \over 2} (is+1)$,
$s = \cot \phi/2$ and so
$t = 1/(1 - e^{i \phi})$. Hence the second equation in (\ref{5.33a}),
together with (\ref{5.62}) in the case $N=1$, gives
(\ref{5.32}). Also, setting $n=0$ in the second equation of
(\ref{5.29}) and equating the right hand side to zero gives the second
initial condition in (\ref{5.in}). The first initial condition in
(\ref{5.in}) follows immediately from the first equation in (\ref{5.29}).
\hfill $\square$

\smallskip

>From the Okamoto theory \cite{Ok87b} we know $p=p(t;\boldsymbol{\alpha}), \,
q=q(t;\boldsymbol{\alpha})$ must satisfy a number of transformation formulas
with respect to $t$ and $\boldsymbol{\alpha}$. Thus with
$$
\boldsymbol{\alpha}^1 
:= (\alpha_0,\alpha_1,\alpha_2,\alpha_4,\alpha_3) \: \: \:
\boldsymbol{\alpha}^2 
:= (\alpha_0,\alpha_4,\alpha_2,\alpha_3,\alpha_1) \: \: \:
\boldsymbol{\alpha}^3 := (\alpha_4,\alpha_1,\alpha_2,\alpha_3,\alpha_0)
$$
one has
\begin{eqnarray}\label{5.34'}
&& p(1-t,\boldsymbol{\alpha}^1) = - p(t;\boldsymbol{\alpha}), \qquad
q(1-t,\boldsymbol{\alpha}^1) = 1 - q(t;\boldsymbol{\alpha}) 
\nonumber\\
&& p \Big ( {1 \over t};\boldsymbol{\alpha}^2 \Big ) = - \alpha_2
q(t;\boldsymbol{\alpha}) - 
q^2(t;\boldsymbol{\alpha}) p(t;\boldsymbol{\alpha}),
\qquad
q({1 \over t};\boldsymbol{\alpha}^2) = 
{1 \over q(t;\boldsymbol{\alpha})} \nonumber \\
&& p\Big ( {t \over t - 1} ; 
\boldsymbol{\alpha}^3 \Big ) = - (t-1) p(t;\boldsymbol{\alpha}),
\qquad
q\Big ( {t \over t - 1};\boldsymbol{\alpha}^3 \Big ) = 
{t - q(t;\boldsymbol{\alpha}) \over
t - 1}.
\end{eqnarray}
Setting $t=1/(1 - e^{i \phi})$ and inverting these formulas we could,
if required, write
down a variant of Proposition \ref{p.8} in each of these cases which
implicitly involves the variables $1-t$, $1/t$ and $t/(t-1)$ respectively,
but at the expense of permuting the $\alpha$'$s$. Consider in particular the
first mapping involving $t \mapsto 1-t$, $\boldsymbol{\alpha} \mapsto
\boldsymbol{\alpha}^1$. With $t = 1/(1 - e^{i \phi})$ and $\alpha_3 = -
w - \mu$, $\alpha_4 = - \bar{w} - \mu$ this corresponds to simply taking the
comlex conjugate. Indeed making the replacements $p \mapsto -P$,
$q \mapsto 1 - Q$ we see that the right hand side of (\ref{5.34}) formally
becomes equal to its complex conjugate, provided we identify $P,Q$ with
$\bar{p}$, $\bar{q}$ respectively. 

Let us now turn our attention to the special case
$2 \mu \in \zz_{\ge 0}$ and $\xi = 0$. Then according to
(\ref{m.1.a}), with $t$ given by (\ref{m.0}) $t^{N\mu} \tau^{VI}[N]$
is a polynomial in $t$,
\begin{eqnarray}\label{5.35}
\lefteqn{
t^{N\mu} \tau^{VI}[N]
= \det \Big [ {\Gamma(2w_1+1) \over
\Gamma(\bar{w}-\mu+j-k+1) \Gamma(w+\mu-j+k+1)}
} \nonumber \\
&& \qquad \qquad 
\times {}_2F_1(-2\mu,-w-\mu+j-k;\bar{w}-\mu+j-k+1;t) \Big ]_{j,k=1,\dots,N}.
\end{eqnarray}
Furthermore in this situation (\ref{5.32}) reduces to
\begin{equation}\label{5.36}
q_0 = 1 - {t \over 2 \mu} {d \over dt} \log {}_2 F_1(-2 \mu,-w-\mu;\bar{w}-
\mu+1;t) = {{}_2 F_1(-2\mu+1,-w-\mu;\bar{w}-\mu+1;t) \over
{}_2 F_1(-2\mu,-w-\mu;\bar{w}-\mu+1;t)}.
\end{equation}
Also
for $2 \mu \in \zz_{\ge 0}$ we have the duality type relation between
averages \cite[eq.~(3.42)]{FW02}
\begin{equation}
\Big \langle \prod_{l=1}^N z_l^{(\eta_1 - \eta_2)/2} |1+z_l|^{\eta_1+
\eta_2}(1 + t z_l)^{2 \mu} \Big \rangle_{U(N)} \propto
\Big \langle \prod_{l=1}^{2 \mu} z_l^{(\eta_1 + 2 \eta_2)/2} |1+z_l|^{\eta_1}
(1 + (1-t) z_l)^{N} \Big \rangle_{U(2 \mu)}.
\end{equation}
Recalling (\ref{4.3}) we thus have
\begin{equation}
\tau^{VI}[N](t;\mu,w_1,w_2;0) \propto
\tau^{VI}[2\mu](1-t;N/2,{\bar{w}-\mu \over 2}, {i \over 2}(\bar{w}
+2w + \mu + N);0),
\end{equation}
which is the PVI analogue of (\ref{4.31}) and (\ref{2.15a}). Setting
$t=0$ shows the proportionality constant to be equal to
$$
{(2 \mu)! \over N!} 
{M_N(-\mu+\bar{w},\mu+w) \over M_{2\mu}(2w_1+N,-\mu-w)}.
$$

Each of the quantities $q_n,p_n,f_n,g_n$ in Proposition \ref{p.8} can
be written as a ratio of $\tau$-functions. Introducing for convenience
\begin{equation}\label{r.el}
\hat{\tau}^{VI}[n](t;\mu,w,\bar{w};\xi) :=
t^{N \mu} \tau^{VI}[N](t;\mu,w_1,w_2;\xi), 
\end{equation}
on the basis
of exact tabulations with initial condition (\ref{5.36}) we are led to
the formulas
\begin{eqnarray}
q_n & = & {\hat{\tau}^{VI}[n+1](t;\mu-1/2,w+1/2,
\bar{w} - 1/2; \xi ) \over
\hat{\tau}^{VI}[n+1](t;\mu,w,\bar{w};\xi) }
{\hat{\tau}^{VI}[n](t;\mu,w,\bar{w}+1;\xi) \over
\hat{\tau}^{VI}[n](t;\mu-1/2,w+1/2,\bar{w}+1/2;\xi)} 
\label{qn.1} \\
p_n & = & 2 \mu (t-1) {\hat{\tau}^{VI}[n+1](t;\mu,w,\bar{w}; \xi ) \over
\hat{\tau}^{VI}[n](t;\mu,w,\bar{w}; \xi )} \nonumber \\
&& \times
{\hat{\tau}^{VI}[n](t;\mu-1/2,w+1/2,
\bar{w} + 1/2; \xi ) \over
\hat{\tau}^{VI}[n](t;\mu,w,\bar{w}+1;\xi) }
{\hat{\tau}^{VI}[n-1](t;\mu+1/2,w+1/2,\bar{w}+1/2;\xi) \over
\hat{\tau}^{VI}[n](t;\mu,w+1,\bar{w};\xi)}
\label{pn.1} \\
f_n & = & - (n+1)t {\hat{\tau}^{VI}[n+1](t;\mu-1/2,w-1/2,
\bar{w} + 1/2; \xi ) \over
\hat{\tau}^{VI}[n+1](t;\mu,w,\bar{w};\xi) }
{\hat{\tau}^{VI}[n](t;\mu,w+1,\bar{w};\xi) \over
\hat{\tau}^{VI}[n](t;\mu-1/2,w+1/2,\bar{w}+1/2;\xi)} \nonumber
\\ \label{fn.1} \\
g_n & = & -{1 \over t} {\hat{\tau}^{VI}[n+1](t;\mu-1/2,w+1/2,
\bar{w} - 1/2; \xi ) \over
\hat{\tau}^{VI}[n+1](t;\mu-1/2,w-1/2,\bar{w}+1/2;\xi) }
{\hat{\tau}^{VI}[n](t;\mu,w,\bar{w}+1;\xi) \over
\hat{\tau}^{VI}[n](t;\mu,w+1,\bar{w};\xi)}.
\label{gn.1}
\end{eqnarray}
We have not completed a proof of these relations. However, as with the
formulas (\ref{4.33}), an outline of how one goes about proving
(\ref{qn.1})--(\ref{gn.1}) is provided by the proof of
Proposition \ref{pr.4}. Here the matter is complicated by there being
four fundamental Schlesinger operators instead of the two in PIII$'$
theory, and the fact that $p_n, q_n$ are functions of $1/(1-t)$ rather
than $t$. Some details of dealing with the first of these complications is
given in \cite{Ma03}, while use of the transformation identities (\ref{5.34'})
is required to deal with the second. Such arguing can only be used to 
establish (\ref{qn.1})--(\ref{gn.1}) up to proportionality constants.
To determine the latter, we proceed on the assumption that the
proportionality is independent of $\xi$, allowing us to set $\xi = 0$.
Then according to (\ref{5.35}) and the formulas of Proposition
\ref{p.8} we must have
$$
q_n \mathop{\sim}\limits_{t \to 0} 1, \quad
g_n \mathop{\sim}\limits_{t \to \infty} - q_n, \quad
f_n \mathop{\sim}\limits_{t \to \infty} - (n+1), \quad
p_n \mathop{\sim}\limits_{t \to 0} {2 \mu n \over \bar{w} - \mu + n}.
$$
On the other hand it follows from (\ref{r.el}) and the definition of
$M_N(a,b)$ in (\ref{m.n}) that with $\xi=0$
$$
\hat{\tau}^{VI}[n] \mathop{\sim}\limits_{t \to 0} {1 \over n!}
M_n(\bar{w}-\mu,w+\mu), \qquad
\hat{\tau}^{VI}[n] \mathop{\sim}\limits_{t \to \infty}
t^{2 n \mu} {1 \over n!} M_n(\bar{w}+\mu,w-\mu).
$$
Now using the evaluation formula in (\ref{m.n}) we deduce the
proportionalities in (\ref{qn.1})--(\ref{gn.1}).

In relation to the PVI $\tau$-function $\tilde{\tau}^{VI}[n]$ as
specified by (\ref{1.8}) we must first recall some theory
from \cite{FW02}. Thus we know that  the working leading to (\ref{5.60})
can be carried through with
$$
\Big ( \int_{-\infty}^\infty - \xi \int_t^\infty \Big ) \mapsto
\Big ( \int_0^1 - \xi \int_t^1 \Big ),
$$
and hence the PVI system admits a $\tau$-function sequence
\begin{eqnarray*}
&& \tau_n^{VI}(t;\mathbf{v}^{(0)}) =
{(-1)^{n(n-1)/2} \over n!} \\
&& \qquad \times
\prod_{j=1}^{n-1}(1+v_3^{(0)}-v_2^{(0)})_j
(1+v_2^{(0)}-v_1^{(0)})_j
\Big ( \int_0^1 - \xi \int_t^1 \Big )du_1 \cdots
\Big ( \int_0^1 - \xi \int_t^1 \Big )du_n \nonumber \\
&& \qquad \times
\prod_{i=1}^n u_i^{-v_2^{(0)}-(v_3^{(0)}+n)}(1 - u_i)^{v_2^{(0)}-(
v_3^{(0)}+n)} (t-u_i)^{-(v_1^{(0)}+v_4^{(0)})}
\prod_{1 \le j < k \le n} (u_k - u_j)^2.
\end{eqnarray*}
Comparison with (\ref{1.8}) shows
\begin{eqnarray}
&& \tau_N^{VI}\Big ( t; ( {1 \over 2}(a+b), {1 \over 2}(b-a),
- {1 \over 2}(a+b) , - {1 \over 2}(a+b)-\mu) \Big ) \nonumber
\\
&& \qquad = (-1)^{N(N-1)/2} {J_N(a-N,b-N) \over N!}
\prod_{j=1}^N (1-a)_j (1-b)_j \tilde{\tau}^{VI}[N]
(t;\mu,a-N,b-N;\xi)
\end{eqnarray}
where
\begin{eqnarray}
J_N(a,b) &:= & \int_0^1 dx_1 \, x_1^a(1-x_1)^b \cdots \int_0^1 dx_N \,
x_N^a(1-x_N)^b \prod_{1 \le j < k \le N}(x_k-x_j)^2 \nonumber \\
& = & \prod_{j=0}^{N-1} {\Gamma(a+1+j) \Gamma(b+1+j) \Gamma(2+j) \over
\Gamma(a+b+1+N+j) }
\end{eqnarray}
It follows from this and (\ref{5.23a}) that we can compute
$\{\tilde{\tau}^{VI}[N](t;\mu,a-N,b-N;\xi) 
\}_{N=0,1,\dots}$ by a recurrence analogous to that
in Proposition \ref{p.8}, although we do not pursue the details.

\section{Applications}
\setcounter{equation}{0}
In this section we will present results from the numerical evaluation
of examples of the $\tau$-functions (\ref{4.1}) and (\ref{4.3}) based
on the recurrences of Propositions \ref{p.3} and \ref{p.8}
respectively. Consider first the $\tau$-function (\ref{4.1}). The cases
$\mu=0$ and $\mu=2$ have particular significance. Thus let
$E_{N,2}(0,(0,s);x^a e^{-x})$ denote the probability that there are
no eigenvalues in the interval $(0,s)$ of the LUE${}_N$ as
specified by the eigenvalue probability density function (\ref{1.9}),
and let $p_{N,2}(0,(0,s);x^ae^{-x})$ denote the probability density
of the smallest eigenvalue in the same ensemble. These two quantities
are inter-related by a single differentiation,
\begin{equation}\label{7.0}
p_{N,2}(0,(0,s);x^ae^{-x}) = - {d \over ds} 
E_{N,2}(0,(0,s);x^a e^{-x}).
\end{equation}

To make contact with (\ref{4.1}) consider the scaled limit of these
quantities,
\begin{eqnarray*}
E_2^{\rm hard}(0,(0,t)) & := & \lim_{N \to \infty}
E_{N,2}(0,(0,t/4N);x^a e^{-x}) \\
p_2^{\rm hard}(0,t) & := & \lim_{N \to \infty} {1 \over 4N}
p_{N,2}(0,t/4N)
\end{eqnarray*}
(the reason for the superscripts ``hard'' is that the neighbourhood of
the origin in the Laguerre ensemble is referred to as the hard edge;
see e.g.~\cite{Fo93a}). Now we know from \cite{FH94} that
\begin{eqnarray}
&& E_2^{\rm hard}(0,(0,t)) \: = \: e^{-t/4} \det [ I_{j-k}(\sqrt{t})
]_{j,k=1,\dots,a} \: = \: e^{-t/4} \tau^{III'}[a](t;0) \label{7.1} \\
&& p_2^{\rm hard}(0,t) \: = \: {1 \over 4} e^{-t/4} \det [ I_{2+j-k}(\sqrt{t})
]_{j,k=1,\dots,a} \: = \: {1 \over 4}
e^{-t/4} \tau^{III'}[a](t;2) \label{7.2} 
\end{eqnarray}
where in both cases the second equality follows from (\ref{11.1}).
Analogous to  (\ref{7.0}) we have
$$
p_2^{\rm hard}(0,t)= - {d \over dt} E_2^{\rm hard}(0,(0,t)).
$$

The large $a$ limit of (\ref{7.1}), (\ref{7.2}) is particularly
interesting. Thus according to the Baik-Deift-Johansson theorem
\cite{BDJ98} (see \cite{BF03} for a recent simplified proof)
\begin{equation}\label{as0}
\lim_{a \to \infty} E_2^{\rm hard}(0,(0,a^2-2a(a/2)^{1/3}s)) =
E_2^{\rm soft}(0,(s,\infty))
\end{equation}
where $E_2^{\rm soft}(0,(s,\infty))$ denotes the scaled probability
of no eigenvalues in the neighbourhood of infinity, and similarly
$$
\lim_{a \to \infty} (2a(a/2)^{1/3}s) p_2(0,a^2-2a(a/2)^{1/3}s)
= p_2^{\rm soft}(0,s)
$$
where $p_2^{\rm soft}(0,s)$ denotes the scaled distribution of the largest
eigenvalue. Let us then address the task of computing
\begin{equation}\label{as}
E_2^{\rm hard}(0,(0,a^2-2a(a/2)^{1/3}s)) =: g^{\rm hard}(a;s)
\end{equation}
using (\ref{7.1}) and the recurrence scheme of Proposition \ref{p.3}.
First it is clear that for large $a$ and $s$ of order unity the
sequence
$$
\{ e^{-t/4} \tau^{III'}[n](t;0) \Big |_{a^2 - 2(a/2)^{1/3} s}
\}_{n=0,1,\dots,a}
$$ 
consists initially of numbers very small in magnitude. Hence it is
necessary to work with high precision arithmetic throughout the
calculation to ensure an accurate final result for 
the final member, which is equal to $g^{\rm hard}(a;s)$. This
sequence in turn is calculated in terms of the sequence $\{P_n\}_{n=0,1,
\dots,a-1}$ as specified by the recurrence in Proposition \ref{p.3},
with $t$ replaced by $t/4$. For the specific value $s=0.5$ the results
of Table \ref{t1} are thereby obtained.

\begin{table}
\begin{center}
\begin{tabular}{|c|c|}\hline
$a$ & $g^{\rm hard}(a;0.5)$ \\[.1cm]\hline
60 & 0.991338737 \\
80 & 0.991201326 \\
100 & 0.991111203 \\
120 & 0.991046762 \\
140 & 0.990997995 \\
160 & 0.990959574 \\ \hline
\end{tabular}
\end{center}
\caption{\label{t1} Tabulation of $g^{\rm hard}(a;0.5)$ as specified by
(\ref{as}) in the case $s=0.5$.}
\end{table}

The data fits well the extrapolation
$$
g^{\rm hard}(a;0.5) = g_0 + {g_1 \over a^{2/3}} + {g_2 \over a}
$$
giving $g_0 = 0.990543$ and thus from (\ref{as0}) predicting
\begin{equation}\label{as8}
E_2^{\rm soft}(0,(0.5,\infty)) = 0.990543.
\end{equation}
In fact $E_2^{\rm soft}(0,(s,\infty))$ is known in terms of a
particular Painlev\'e II transcendent $q(s)$ \cite{TW94a}.
High precision data by way of the values of
$E_2^{\rm soft}(0,(0,\infty))$, $q(0), q'(0)$ to 50 decimals have
recently been given \cite{PS02}, allowing for accurate determination of
$E_2^{\rm soft}$ for general $s$. 
One finds $E_2^{\rm soft}(0,(0.5,\infty)) = 0.990544...$,
showing us that (\ref{as8}) is accurate to 1 part in $10^6$.

We now turn our attention to a particular example of the $\tau$-function
(\ref{4.3}). Let $p_{N-2,0}^{\rm CUE}(\theta)$ denote the probability
density function for the spacing between consecutive eigenvalues in the
CUE${}_N$ or equivalently $U(N)$. Then as noted in \cite{FW02}, it
follows from the definitions that
\begin{equation}
\Big ( {2 \pi \over N} \Big ) p_{N-2}^{\rm CUE}\Big ( 
{2 \pi X \over N} \Big )
= {1 \over 3} (N^2 - 1) \sin^2 {\pi X \over N} \,
{\tau^{VI}[N-2](e^{2 \pi i X/N};1,1,0;1) \over
\tau^{VI}[N-2](1;1,1,0;1) }.
\end{equation}
Use of (\ref{m.n}) shows
$$
\tau^{VI}[N](1;1,1,0;1) = {(N+2)^2(N+1)(N+3) \over 12}
$$
and thus
\begin{equation}\label{tt1.0}
\Big ( {2 \pi \over N} \Big ) 
p_{N-2}^{\rm CUE}\Big ( {2 \pi X \over N} \Big )
= {4 \over N^2} \sin^2 {\pi X \over N}
\, \tau^{VI}[N-2](e^{2 \pi i X/N};1,1,0;1).
\end{equation}
According to Proposition \ref{p.8} the key quantity in computing
$\{\tau^{VI}[n]\}$ by recurrence is $\tau^{VI}[1]$. Now, from
(\ref{m.1.a}) and (\ref{m.2.a})
\begin{eqnarray}\label{tt1.1}
\lefteqn{\tau^{VI}(e^{i \phi};1,1,0;1)} \nonumber \\
&&= e^{-i\phi} {}_2 F_1(-2,-2;1;e^{i \phi}) +
{1 \over 60 \pi i } e^{-i \phi} (1 - e^{i \phi})^5
{}_2 F_1(3,3;6;1-e^{i \phi})
\end{eqnarray}
(identities for the ${}_2 F_1$ function can be used to check that
this quantity is real. Using this in  Proposition \ref{p.8}, and again
using high precision computing, for the specific value $X=1/10$ and a
sequence of $N$ values, we
evaluated (\ref{tt1.0}), obtaining the data listed in Table \ref{t2}.

\begin{table}
\begin{center}
\begin{tabular}{|c|c|}\hline
$N$ & ${2 \pi \over N} p_{N-2}^{\rm CUE}( {2 \pi X \over N})
\Big |_{X=1/10}$ \\[.1cm]\hline
10 & 0.03215040321 \\
30 & 0.03243339939 \\
50 & 0.03245603495 \\
70 & 0.03246227118 \\
90 & 0.03246483751 \\ \hline
\end{tabular}
\end{center}
\caption{\label{t2} Tabulation of the scaled probability density at
$X=0.1$ for the spacing between consecutive eigenvalues in the CUE${}_N$.}
\end{table}

The limiting distribution
\begin{equation}
p_2^{\rm bulk}(X) = \lim_{N \to \infty} {2 \pi \over N}
p_{N-2}^{\rm CUE}( {2 \pi X \over N} )
\end{equation}
can itself be expressed in terms of a Painlev\'e transcendent
\cite{JMMS80,FW00e,FW02}. Moreover its power series about $X=0$ is known
to high accuracy \cite{Gr02}, and from this we can compute
\begin{equation}\label{gr.1}
p_2^{\rm bulk}(X) \Big |_{X=1/10} =
0.032468767196387...
\end{equation}
Extrapolating the data of Table \ref{t2} using the ansatz
$$
{2 \pi \over N}
p_{N-2}^{\rm CUE}( {2 \pi X \over N} ) = s_0 + {s_1 \over N^2} +
{s_2 \over N^4}
$$
gives $s_0=0.032468767193...$ which agrees with (\ref{gr.1}) to 3 parts in
$10^{12}$.

\section*{Acknowledgements}
This work was supported by the Australian Research Council.

\bibliographystyle{plain}
\bibliography{book1}

\begin{thebibliography}{10}

\bibitem{AvM02}
M.~Adler and P.~van Moerbeke.
\newblock Recursion relations for unitary integrals, combinatorics and the
  {T}oeplitz lattice.
\newblock arXiv:math-ph/0201063.

\bibitem{Av95}
M.~Adler and P.~van Moerbeke.
\newblock Matrix integrals, {Toda} symmetries, {Virasora} constraints and
  orthogonal polynomials.
\newblock {\em Duke Math. Journal}, 80:863--911, 1995.

\bibitem{Ba02}
J.~Baik.
\newblock {Painlev\'e} expressions for {LOE}, {LSE}, and interpolating
  ensembles.
\newblock {\em Int. Math. Res. Notices}, (33):1739--1789, 2002.

\bibitem{BDJ98}
J.~Baik, P.~Dieft, and K.~Johansson.
\newblock On the distribution of the length of the longest increasing
  subsequence of random permutations.
\newblock {\em J. Amer. Math. Soc.}, 12:1119--1178, 1999.

\bibitem{Bo02}
A.~Borodin.
\newblock Discrete gap probabilities and discrete {P}ainlev\'e equations.
\newblock arXiv:math-ph/0111008.

\bibitem{BB02}
A.~Borodin and D. Boyarchenko.
\newblock Distribution of the first particle in discrete orthogonal polynomial
  ensembles.
\newblock arXiv:math-ph/0204001.

\bibitem{Bo02a}
A.~Borodin.
\newblock Isomonodromy transformations of linear systems of difference
  equations.
\newblock arXiv:math.CA/0209144.

\bibitem{Bo00}
A.~Borodin.
\newblock Riemann-{H}ilbert problem and the discrete {Bessel} kernel.
\newblock {\em Inter. Math. Res. Notices}, no. 9:467--494, 2000.

\bibitem{BF03}
A.~Borodin and P.J. Forrester.
\newblock Increasing subsequences and the hard-to-soft transition in matrix
  ensembles.
\newblock {\em J.Phys. A}, 36:2963--2981, 2003.

\bibitem{BO01}
A.~Borodin and G.~Olshanski.
\newblock Infinite random matrices and ergodic measures.
\newblock {\em Commun. Math. Phys.}, 223:87--123, 2001.

\bibitem{Fo02}
P.J. Forrester.
\newblock Log-gases and {Random} {Matrices}.
\newblock www.ms.unimelb.edu.au/\~{}matpjf/matpjf.html.

\bibitem{Fo93a}
P.J. Forrester.
\newblock The spectrum edge of random matrix ensembles.
\newblock {\em Nucl. Phys. B}, 402:709--728, 1993.

\bibitem{FH94}
P.J. Forrester and T.D. Hughes.
\newblock Complex {Wishart} matrices and conductance in mesoscopic systems:
  exact results.
\newblock {\em J. Math. Phys.}, 35:6736--6747, 1994.


\bibitem{FW00e}
P.J. Forrester and N.S. Witte.
\newblock Exact {W}igner surmise type evaluation of the spacing distribution in
  the bulk of the scaled random matrix ensembles.
\newblock {\em Lett. Math. Phys.}, 53:195--200, 2000.

\bibitem{FW00}
P.J. Forrester and N.S. Witte.
\newblock Application of the $\tau$-function theory of {Painlev\'e} equations
  to random matrices: {PIV}, {PII} and the {GUE}.
\newblock {\em Commun. Math. Phys.}, 219:357--398, 2001.


\bibitem{FW01a}
P.J. Forrester and N.S. Witte.
\newblock Application of the $\tau$-function theory of {Painlev\'e} equations
  to random matrices: {PV}, {PIII}, the {LUE}, {JUE} and {CUE}.
\newblock {\em Commun. Pure Appl. Math.}, 55:679--727, 2002.

\bibitem{FW02}
P.J. Forrester and N.S. Witte.
\newblock Application of the $\tau$-function theory of {Painlev\'e} equations
  to random matrices: {PVI}, the {JUE},{CyUE}, {cJUE} and scaled limits.
\newblock math-ph/0204008, 2002.

\bibitem{FW02c}
P.J. Forrester and N.S. Witte.
\newblock $\tau$-function evaluation of gap probabilities in orthogonal and
  symplectic matrix ensembles.
\newblock {\em Nonlinearity}, 15:937--954, 2002.

\bibitem{Gr02}
U.~Grimm.
\newblock Series expansions for level-spacing distributions of the {G}aussian
  unitary random matrix ensemble.
\newblock arXiv:cond-mat/0211279.

\bibitem{JMMS80}
M.~Jimbo, T.~Miwa, Y.~M\^ori, and M.~Sato.
\newblock Density matrix of an impenetrable {Bose} gas and the fifth
  {Painlev\'e} transcendent.
\newblock {\em Physica}, 1D:80--158, 1980.

\bibitem{K99}
K.~Kajiwara, T.~Masuda, M.~Noumi, Y.~Ohta, and Y.~Yamada.
\newblock Determinant formulas for the {Toda} and discrete {Toda} equations.
\newblock {\em Funkcialaj Ekvacioj}, 44:291--307, 2001.

\bibitem{Ma22}
J.~Malmquist.
\newblock Sur les \'equations diff\'erentialles du second ordre dont
  l'int\'egrale g\'en\'erale a ses points critiques fixes.
\newblock {\em Arkiv Mat. Astron. Fys.}, 18:1--89, 1922.

\bibitem{Ma03}
T.~Masuda.
\newblock Classical transcendental solutions of the painlev\'e equations and
  their degeneration.
\newblock arXiv:nlin.SI/0302026, 2003.

\bibitem{NY99}
M.~Noumi and Y.~Yamada.
\newblock Symmetries in the fourth {Painlev\'e} equation and {O}kamoto
  polynomials. 
\newblock {\em Nagoya Math. J.}, 153:53--86, 1999.

\bibitem{Ok86}
K.~Okamoto.
\newblock Studies of the {Painlev\'e} equations. {III}. {Second} and fourth
  {Painlev\'e} equations, {$P_{II}$} and {$P_{IV}$}.
\newblock {\em Math. Ann.}, 275:221--255, 1986.

\bibitem{Ok87b}
K.~Okamoto.
\newblock Studies of the {Painlev\'e} equations. {I}. {Sixth} {Painlev\'e}
  equation {$P_{VI}$}.
\newblock {\em Ann. Math. Pura Appl.}, 146:337--381, 1987.

\bibitem{Ok87}
K.~Okamoto.
\newblock Studies of the {Painlev\'e} equations. {II}. {Fifth} {Painlev\'e}
  equation {$P_{V}$}.
\newblock {\em Japan J. Math.}, 13:47--76, 1987.

\bibitem{Ok87a}
K.~Okamoto.
\newblock Studies of the {Painlev\'e} equations. {IV}. {Third} {Painlev\'e}
  equation {$P_{III}$}.
\newblock {\em Funkcialaj Ekvacioj}, 30:305--332, 1987.

\bibitem{PS02}
M.~Pr\"ahofer and H.~Spohn.
\newblock Exact scaling functions for one-dimensional stationary {KPZ} growth.
\newblock arXiv:cond-mat/0212519, 2002.

\bibitem{Sa01}
H.~Sakai.
\newblock Rational surfaces associated with affine root systems and geometry of
  the painlev\'e equations.
\newblock {\em Commun. Math. Phys.}, 220:165--229, 2001.

\bibitem{TW94a}
C.A. Tracy and H.~Widom.
\newblock Level-spacing distributions and the {Airy} kernel.
\newblock {\em Commun. Math. Phys.}, 159:151--174, 1994.

\bibitem{Wa98}
H.~Watanabe.
\newblock Defining variety and birational canonical transformations of the
  fifth {Painlev\'e} equation.
\newblock {\em Analysis}, 18:351--357, 1998.

\bibitem{WF00a}
N.S. Witte and P.J. Forrester.
\newblock Gap probabilities in the finite and scaled {Cauchy} random matrix
  ensembles.
\newblock {\em Nonlinearity}, 13:1965--1986, 2000.

\end{thebibliography}

\end{document}